\documentclass[12pt]{article}
\usepackage{amsmath}
\usepackage{multirow}
\usepackage{citesort}
\usepackage{axodraw}
\usepackage{ifthen}

\def\xdim{40}   % Size of the picture-environment
\def\ydim{40}   % Default (without border): 40pt = 14.1mm
\def\xoff{0}    % Position of graphic in the picture-environment
\def\yoff{0}
\def\dash{2}    % Dash size for massless lines
\def\vtxsz{1.2} % Size of vertex
\def\dotsz{1.5} % Size of Dot

\SetScale{1}    % Scale factor of graphic
\SetWidth{0.7}  % Line width

\newcommand{\momsz}{\small} % Textsize in diagrams

\newcommand{\LAHA}[1]{\ifthenelse{\equal{#1}{1}}{\CArc(20,20)(20,90,270)}{\DashCArc(20,20)(20,90,270){\dash}}}
\newcommand{\LAHB}[1]{\ifthenelse{\equal{#1}{1}}{\CArc(20,20)(20,-90,90)}{\DashCArc(20,20)(20,-90,90){\dash}}}
\newcommand{\LAGA}[1]{\ifthenelse{\equal{#1}{1}}{\CArc(35,20)(25,126.87,233.13)}{\DashCArc(35,20)(25,126.87,233.13){\dash}}}
\newcommand{\LAGB}[1]{\ifthenelse{\equal{#1}{1}}{\CArc(5,20)(25,-53.13,53.13)}{\DashCArc(5,20)(25,-53.13,53.13){\dash}}}
\newcommand{\LADA}[1]{\ifthenelse{\equal{#1}{1}}{\CArc(20,20)(20,30,150)}{\DashCArc(20,20)(20,30,150){\dash}}}
\newcommand{\LADB}[1]{\ifthenelse{\equal{#1}{1}}{\CArc(20,20)(20,150,270)}{\DashCArc(20,20)(20,150,270){\dash}}}
\newcommand{\LADC}[1]{\ifthenelse{\equal{#1}{1}}{\CArc(20,20)(20,-90,30)}{\DashCArc(20,20)(20,-90,30){\dash}}}
\newcommand{\LAVA}[1]{\ifthenelse{\equal{#1}{1}}{\CArc(20,20)(20,45,135)}{\DashCArc(20,20)(20,45,135){\dash}}}
\newcommand{\LAVB}[1]{\ifthenelse{\equal{#1}{1}}{\CArc(20,20)(20,135,225)}{\DashCArc(20,20)(20,135,225){\dash}}}
\newcommand{\LAVC}[1]{\ifthenelse{\equal{#1}{1}}{\CArc(20,20)(20,225,315)}{\DashCArc(20,20)(20,225,315){\dash}}}
\newcommand{\LAVD}[1]{\ifthenelse{\equal{#1}{1}}{\CArc(20,20)(20,-45,45)}{\DashCArc(20,20)(20,-45,45){\dash}}}
\newcommand{\LASA}[1]{\ifthenelse{\equal{#1}{1}}{\CArc(20,20)(20,90,150)}{\DashCArc(20,20)(20,90,150){\dash}}}
\newcommand{\LASB}[1]{\ifthenelse{\equal{#1}{1}}{\CArc(20,20)(20,150,210)}{\DashCArc(20,20)(20,150,210){\dash}}}
\newcommand{\LASC}[1]{\ifthenelse{\equal{#1}{1}}{\CArc(20,20)(20,210,270)}{\DashCArc(20,20)(20,210,270){\dash}}}
\newcommand{\LASD}[1]{\ifthenelse{\equal{#1}{1}}{\CArc(20,20)(20,270,330)}{\DashCArc(20,20)(20,270,330){\dash}}}
\newcommand{\LASE}[1]{\ifthenelse{\equal{#1}{1}}{\CArc(20,20)(20,-30,30)}{\DashCArc(20,20)(20,-30,30){\dash}}}
\newcommand{\LASF}[1]{\ifthenelse{\equal{#1}{1}}{\CArc(20,20)(20,30,90)}{\DashCArc(20,20)(20,30,90){\dash}}}
\newcommand{\LAWA}[1]{\ifthenelse{\equal{#1}{1}}{\CArc(20,20)(20,90,180)}{\DashCArc(20,20)(20,90,180){\dash}}}
\newcommand{\LAWB}[1]{\ifthenelse{\equal{#1}{1}}{\CArc(20,20)(20,180,270)}{\DashCArc(20,20)(20,180,270){\dash}}}
\newcommand{\LAWC}[1]{\ifthenelse{\equal{#1}{1}}{\CArc(20,20)(20,-90,0)}{\DashCArc(20,20)(20,-90,0){\dash}}}
\newcommand{\LAWD}[1]{\ifthenelse{\equal{#1}{1}}{\CArc(20,20)(20,0,90)}{\DashCArc(20,20)(20,0,90){\dash}}}
\newcommand{\LAXA}[1]{\ifthenelse{\equal{#1}{1}}{\CArc(20,20)(20,70.53,109.47)}{\DashCArc(20,20)(20,70.53,109.47){\dash}}}
\newcommand{\LAXB}[1]{\ifthenelse{\equal{#1}{1}}{\CArc(20,20)(20,109.47,250.53)}{\DashCArc(20,20)(20,109.47,250.53){\dash}}}
\newcommand{\LAXC}[1]{\ifthenelse{\equal{#1}{1}}{\CArc(20,20)(20,250.53,289.47)}{\DashCArc(20,20)(20,250.53,289.47){\dash}}}
\newcommand{\LAXD}[1]{\ifthenelse{\equal{#1}{1}}{\CArc(20,20)(20,-70.53,70.53)}{\DashCArc(20,20)(20,-70.53,70.53){\dash}}}
\newcommand{\LAZA}[1]{\ifthenelse{\equal{#1}{1}}{\CArc(4.84,28.75)(32.5,-62.20,2.20)}{\DashCArc(4.84,28.75)(32.5,-62.20,2.20){\dash}}}
\newcommand{\LCOO}[1]{\ifthenelse{\equal{#1}{1}}{\CArc(20,35)(5,0,360)}{\DashCArc(20,35)(5,0,360){\dash}}}
\newcommand{\LCOP}[1]{\ifthenelse{\equal{#1}{1}}{\CArc(20,15)(15,90,270)}{\DashCArc(20,15)(15,90,270){\dash}}}
\newcommand{\LCOQ}[1]{\ifthenelse{\equal{#1}{1}}{\CArc(40,15)(25,143.13,216.87)}{\DashCArc(40,15)(25,143.13,216.87){\dash}}}
\newcommand{\LCOR}[1]{\ifthenelse{\equal{#1}{1}}{\CArc(0,15)(25,-36.87,36.87)}{\DashCArc(0,15)(25,-36.87,36.87){\dash}}}
\newcommand{\LCOS}[1]{\ifthenelse{\equal{#1}{1}}{\CArc(20,15)(15,-90,180)}{\DashCArc(20,15)(15,-90,180){\dash}}}
\newcommand{\LCOT}[1]{\ifthenelse{\equal{#1}{1}}{\CArc(20,15)(15,90,210)}{\DashCArc(20,15)(15,90,210){\dash}}}
\newcommand{\LCTU}[1]{\ifthenelse{\equal{#1}{1}}{\CArc(20,15)(15,210,330)}{\DashCArc(20,15)(15,210,330){\dash}}}
\newcommand{\LCUO}[1]{\ifthenelse{\equal{#1}{1}}{\CArc(20,15)(15,-30,90)}{\DashCArc(20,15)(15,-30,90){\dash}}}
\newcommand{\LCOV}[1]{\ifthenelse{\equal{#1}{1}}{\CArc(20,20)(10,90,270)}{\DashCArc(20,20)(10,90,270){\dash}}}
\newcommand{\LCOW}[1]{\ifthenelse{\equal{#1}{1}}{\CArc(20,20)(10,-90,90)}{\DashCArc(20,20)(10,-90,90){\dash}}}
\newcommand{\LCVV}[1]{\ifthenelse{\equal{#1}{1}}{\CArc(20,5)(5,0,360)}{\DashCArc(20,5)(5,0,360){\dash}}}
\newcommand{\LCWD}[1]{\ifthenelse{\equal{#1}{1}}{\CArc(20,30)(10,90,270)}{\DashCArc(20,30)(10,90,270){\dash}}}
\newcommand{\LCWC}[1]{\ifthenelse{\equal{#1}{1}}{\CArc(20,10)(10,90,270)}{\DashCArc(20,10)(10,90,270){\dash}}}
\newcommand{\LCDW}[1]{\ifthenelse{\equal{#1}{1}}{\CArc(20,30)(10,-90,90)}{\DashCArc(20,30)(10,-90,90){\dash}}}
\newcommand{\LCCW}[1]{\ifthenelse{\equal{#1}{1}}{\CArc(20,10)(10,-90,90)}{\DashCArc(20,10)(10,-90,90){\dash}}}

\newcommand{\LLHA}[1]{\ifthenelse{\equal{#1}{1}}{\Line(20,0)(20,40)}{\DashLine(20,0)(20,40){\dash}}}
\newcommand{\LLHB}[1]{\ifthenelse{\equal{#1}{1}}{\Line(2.68,30)(37.32,10)}{\DashLine(2.68,30)(37.32,10){\dash}}}
\newcommand{\LLHC}[1]{\ifthenelse{\equal{#1}{1}}{\Line(2.68,10)(37.32,30)}{\DashLine(2.68,10)(37.32,30){\dash}}}
\newcommand{\LLDA}[1]{\ifthenelse{\equal{#1}{1}}{\Line(2.68,30)(37.32,30)}{\DashLine(2.68,30)(37.32,30){\dash}}}
\newcommand{\LLDB}[1]{\ifthenelse{\equal{#1}{1}}{\Line(2.68,30)(20,0)}{\DashLine(2.68,30)(20,0){\dash}}}
\newcommand{\LLDC}[1]{\ifthenelse{\equal{#1}{1}}{\Line(20,0)(37.32,30)}{\DashLine(20,0)(37.32,30){\dash}}}
\newcommand{\LLEA}[1]{\ifthenelse{\equal{#1}{1}}{\Line(2.68,30)(20,20)}{\DashLine(2.68,30)(20,20){\dash}}}
\newcommand{\LLEB}[1]{\ifthenelse{\equal{#1}{1}}{\Line(20,0)(20,20)}{\DashLine(20,0)(20,20){\dash}}}
\newcommand{\LLEC}[1]{\ifthenelse{\equal{#1}{1}}{\Line(37.32,30)(20,20)}{\DashLine(37.32,30)(20,20){\dash}}}
\newcommand{\LLVA}[1]{\ifthenelse{\equal{#1}{1}}{\Line(5.86,34.14)(5.86,5.86)}{\DashLine(5.86,34.14)(5.86,5.86){\dash}}}
\newcommand{\LLVB}[1]{\ifthenelse{\equal{#1}{1}}{\Line(5.86,5.86)(34.14,5.86)}{\DashLine(5.86,5.86)(34.14,5.86){\dash}}}
\newcommand{\LLVC}[1]{\ifthenelse{\equal{#1}{1}}{\Line(34.14,5.86)(34.14,34.14)}{\DashLine(34.14,5.86)(34.14,34.14){\dash}}}
\newcommand{\LLSA}[1]{\ifthenelse{\equal{#1}{1}}{\Line(28.66,15)(37.32,30)}{\DashLine(28.66,15)(37.32,30){\dash}}}
\newcommand{\LLSB}[1]{\ifthenelse{\equal{#1}{1}}{\Line(28.66,15)(20,0)}{\DashLine(28.66,15)(20,0){\dash}}}
\newcommand{\LLSC}[1]{\ifthenelse{\equal{#1}{1}}{\Line(28.66,15)(37.32,10)}{\DashLine(28.66,15)(37.32,10){\dash}}}
\newcommand{\LLWA}[1]{\ifthenelse{\equal{#1}{1}}{\Line(20,20)(20,40)}{\DashLine(20,20)(20,40){\dash}}}
\newcommand{\LLWB}[1]{\ifthenelse{\equal{#1}{1}}{\Line(20,20)(0,20)}{\DashLine(20,20)(0,20){\dash}}}
\newcommand{\LLWC}[1]{\ifthenelse{\equal{#1}{1}}{\Line(20,20)(20,0)}{\DashLine(20,20)(20,0){\dash}}}
\newcommand{\LLWD}[1]{\ifthenelse{\equal{#1}{1}}{\Line(20,20)(40,20)}{\DashLine(20,20)(40,20){\dash}}}
\newcommand{\LLXA}[1]{\ifthenelse{\equal{#1}{1}}{\Line(20,20)(5.86,34.14)}{\DashLine(20,20)(5.86,34.14){\dash}}}
\newcommand{\LLXB}[1]{\ifthenelse{\equal{#1}{1}}{\Line(20,20)(5.86,5.86)}{\DashLine(20,20)(5.86,5.86){\dash}}}
\newcommand{\LLXC}[1]{\ifthenelse{\equal{#1}{1}}{\Line(20,20)(34.14,5.86)}{\DashLine(20,20)(34.14,5.86){\dash}}}
\newcommand{\LLXD}[1]{\ifthenelse{\equal{#1}{1}}{\Line(20,20)(34.14,34.14)}{\DashLine(20,20)(34.14,34.14){\dash}}}
\newcommand{\LLIA}[1]{\ifthenelse{\equal{#1}{1}}{\Line(13.33,20)(13.33,1.14)}{\DashLine(13.33,20)(13.33,1.14){\dash}}}
\newcommand{\LLIB}[1]{\ifthenelse{\equal{#1}{1}}{\Line(26.67,20)(26.67,1.14)}{\DashLine(26.67,20)(26.67,1.14){\dash}}}
\newcommand{\LLIC}[1]{\ifthenelse{\equal{#1}{1}}{\Line(13.33,20)(26.67,20)}{\DashLine(13.33,20)(26.67,20){\dash}}}
\newcommand{\LLID}[1]{\ifthenelse{\equal{#1}{1}}{\Line(13.33,20)(13.33,38.86)}{\DashLine(13.33,20)(13.33,38.86){\dash}}}
\newcommand{\LLIE}[1]{\ifthenelse{\equal{#1}{1}}{\Line(26.67,20)(26.67,38.86)}{\DashLine(26.67,20)(26.67,38.86){\dash}}}
\newcommand{\LLIF}[1]{\ifthenelse{\equal{#1}{1}}{\Line(26.67,20)(13.33,38.86)}{\DashLine(26.67,20)(13.33,38.86){\dash}}}
\newcommand{\LLIG}[1]{\ifthenelse{\equal{#1}{1}}{\Line(13.33,20)(26.67,38.86)}{\DashLine(13.33,20)(26.67,38.86){\dash}}}
\newcommand{\LLCO}[1]{\ifthenelse{\equal{#1}{1}}{\Line(20,15)(20,30)}{\DashLine(20,15)(20,30){\dash}}}
\newcommand{\LLCT}[1]{\ifthenelse{\equal{#1}{1}}{\Line(20,15)(7.01,7.5)}{\DashLine(20,15)(7.01,7.5){\dash}}}
\newcommand{\LLCU}[1]{\ifthenelse{\equal{#1}{1}}{\Line(20,15)(32.99,7.5)}{\DashLine(20,15)(32.99,7.5){\dash}}}
\newcommand{\LLOV}[1]{\ifthenelse{\equal{#1}{1}}{\Line(20,10)(20,30)}{\DashLine(20,10)(20,30){\dash}}}
\newcommand{\LLOT}[1]{\ifthenelse{\equal{#1}{1}}{\Line(7.01,7.5)(20,30)}{\DashLine(7.01,7.5)(20,30){\dash}}}
\newcommand{\LLUO}[1]{\ifthenelse{\equal{#1}{1}}{\Line(32.99,7.5)(20,30)}{\DashLine(32.99,7.5)(20,30){\dash}}}

\newcommand{\LVAA}{\Vertex(20,40){\vtxsz}}
\newcommand{\LVAB}{\Vertex(13.33,38.86){\vtxsz}}
\newcommand{\LVAC}{\Vertex(5.86,34.14){\vtxsz}}
\newcommand{\LVAD}{\Vertex(2.68,30){\vtxsz}}
\newcommand{\LVAE}{\Vertex(0,20){\vtxsz}}
\newcommand{\LVAF}{\Vertex(5.86,5.86){\vtxsz}}
\newcommand{\LVAG}{\Vertex(13.33,1.14){\vtxsz}}
\newcommand{\LVAH}{\Vertex(20,0){\vtxsz}}
\newcommand{\LVAI}{\Vertex(26.67,1.14){\vtxsz}}
\newcommand{\LVAJ}{\Vertex(34.14,5.86){\vtxsz}}
\newcommand{\LVAK}{\Vertex(37.32,10){\vtxsz}}
\newcommand{\LVAL}{\Vertex(40,20){\vtxsz}}
\newcommand{\LVAM}{\Vertex(37.32,30){\vtxsz}}
\newcommand{\LVAN}{\Vertex(34.14,34.14){\vtxsz}}
\newcommand{\LVAO}{\Vertex(26.67,38.86){\vtxsz}}
\newcommand{\LVMA}{\Vertex(13.33,20){\vtxsz}}
\newcommand{\LVMB}{\Vertex(20,20){\vtxsz}}
\newcommand{\LVMC}{\Vertex(26.67,20){\vtxsz}}
\newcommand{\LVZA}{\Vertex(28.66,15){\vtxsz}}
\newcommand{\LVSA}{\Vertex(2.68,10){\vtxsz}}
\newcommand{\LVOO}{\Vertex(20,30){\vtxsz}}
\newcommand{\LVOC}{\Vertex(20,15){\vtxsz}}
\newcommand{\LVOT}{\Vertex(7.01,7.5){\vtxsz}}
\newcommand{\LVOU}{\Vertex(32.99,7.5){\vtxsz}}
\newcommand{\LVVV}{\Vertex(20,10){\vtxsz}}

\newcommand{\MFA}[5]{\LAHA{#1}\LAGA{#2}\LLHA{#3}\LAGB{#4}\LAHB{#5}\LVAH\LVAA}
\newcommand{\MFB}[5]{\LCOO{#1}\LCOP{#2}\LCOQ{#3}\LCOR{#4}\LCOS{#5}\LVAH\LVOO}
\newcommand{\MFC}[5]{\LCOO{#1}\LCOV{#2}\LLOV{#3}\LCOW{#4}\LCVV{#5}\LVVV\LVOO}
\newcommand{\MGA}[6]{\LADA{#1}\LADB{#2}\LLDB{#3}\LLDC{#4}\LAZA{#5}\LADC{#6}\LVAD\LVAH\LVAM}
\newcommand{\MGB}[6]{\LADA{#1}\LLDA{#2}\LADB{#3}\LLDB{#4}\LLDC{#5}\LADC{#6}\LVAD\LVAH\LVAM}
\newcommand{\MGC}[6]{\LCOO{#1}\LCOT{#2}\LLOT{#3}\LLUO{#4}\LCUO{#5}\LCTU{#6}\LVOT\LVOU\LVOO}
\newcommand{\MGD}[6]{\LCWD{#1}\LLWA{#2}\LCDW{#3}\LCWC{#4}\LLWC{#5}\LCCW{#6}\LVAH\LVAA\LVMB}
\newcommand{\MHA}[7]{\LAVA{#1}\LAVB{#2}\LLVA{#3}\LLVC{#4}\LAVD{#5}\LLVB{#6}\LAVC{#7}\LVAC\LVAF\LVAJ\LVAN}
\newcommand{\MHB}[7]{\LADA{#1}\LLDA{#2}\LADB{#3}\LLEA{#4}\LLEB{#5}\LLEC{#6}\LADC{#7}\LVAD\LVAH\LVAM\LVMB}
\newcommand{\MHC}[7]{\LCOO{#1}\LCOT{#2}\LLCO{#3}\LCUO{#4}\LLCT{#5}\LLCU{#6}\LCTU{#7}\LVOO\LVOC\LVOT\LVOU}
\newcommand{\MIA}[8]{\LADA{#1}\LADB{#2}\LLDB{#3}\LLSB{#4}\LASD{#5}\LLSC{#6}\LLSA{#7}\LASE{#8}\LVAD\LVAH\LVAM\LVAK\LVZA}
\newcommand{\MIB}[8]{\LAWA{#1}\LLWB{#2}\LLWA{#3}\LLWD{#4}\LAWD{#5}\LAWB{#6}\LLWC{#7}\LAWC{#8}\LVAA\LVAE\LVAH\LVAL\LVMB}
\newcommand{\MJB}[9]{\LAXB{#1}\LLID{#2}\LAXA{#3}\LLIE{#4}\LAXD{#5}\LLIA{#6}\LLIC{#7}\LLIB{#8}\LAXC{#9}\LVAB\LVAG\LVAI\LVAO\LVMA\LVMC}
\newcommand{\MJC}[9]{\LASB{#1}\LLHB{#2}\LASA{#3}\LLHA{#4}\LASF{#5}\LLHC{#6}\LASE{#7}\LASC{#8}\LASD{#9}\LVAD\LVAH\LVAM\LVAA\LVAK\LVSA}

\newcommand{\Topo}[1]{
\ifthenelse{\equal{#1}{41501}}{\MFA{1}{0}{0}{0}{0}}{}
\ifthenelse{\equal{#1}{41502}}{\MFB{1}{1}{0}{0}{0}}{}
\ifthenelse{\equal{#1}{41503}}{\MFC{1}{0}{1}{0}{1}}{}
\ifthenelse{\equal{#1}{41504}}{\MFA{1}{1}{1}{0}{0}}{}
\ifthenelse{\equal{#1}{4150401x2}}{\MFA{1}{1}{1}{0}{0}\Vertex(30,20){\dotsz}}{}
\ifthenelse{\equal{#1}{4150403x3}}{\MFA{1}{1}{1}{0}{0}\Vertex(20,13.33){\dotsz}\Vertex(20,26.67){\dotsz}}{}
\ifthenelse{\equal{#1}{41505}}{\MFB{1}{1}{1}{0}{1}}{}
\ifthenelse{\equal{#1}{41603}}{\MGA{1}{0}{0}{1}{0}{0}}{}
\ifthenelse{\equal{#1}{41604}}{\MGC{1}{0}{0}{0}{0}{1}}{}
\ifthenelse{\equal{#1}{41605}}{\MGD{0}{1}{0}{0}{1}{0}}{}
\ifthenelse{\equal{#1}{416010}}{\MGB{1}{0}{0}{0}{0}{0}}{}
\ifthenelse{\equal{#1}{416011}}{\MGB{0}{0}{1}{0}{0}{1}}{}
\ifthenelse{\equal{#1}{416017}}{\MGA{1}{0}{0}{1}{1}{1}}{}
\ifthenelse{\equal{#1}{416018}}{\MGA{1}{0}{0}{0}{0}{0}}{}
\ifthenelse{\equal{#1}{416021}}{\MGA{1}{0}{0}{0}{1}{1}}{}
\ifthenelse{\equal{#1}{416022}}{\MGA{0}{0}{1}{1}{1}{1}}{}
\ifthenelse{\equal{#1}{416023}}{\MGB{1}{0}{1}{1}{1}{1}}{}
\ifthenelse{\equal{#1}{41602303x2}}{\MGB{1}{0}{1}{1}{1}{1}\Vertex(11.34,15){\dotsz}}{}
\ifthenelse{\equal{#1}{416024}}{\MGB{1}{0}{1}{1}{0}{0}}{}
\ifthenelse{\equal{#1}{41602403x2}}{\MGB{1}{0}{1}{1}{0}{0}\Vertex(11.34,15){\dotsz}}{}
\ifthenelse{\equal{#1}{416025}}{\MGB{1}{1}{1}{0}{0}{1}}{}
\ifthenelse{\equal{#1}{41602503x2}}{\MGB{1}{1}{1}{0}{0}{1}\Vertex(20,30){\dotsz}}{}
\ifthenelse{\equal{#1}{417019}}{\MHA{1}{0}{0}{0}{0}{0}{0}}{}
\ifthenelse{\equal{#1}{417030}}{\MHB{0}{0}{1}{0}{1}{1}{0}}{}
\ifthenelse{\equal{#1}{417036}}{\MHB{0}{1}{0}{1}{1}{0}{0}}{}
\ifthenelse{\equal{#1}{417038}}{\MHC{1}{1}{0}{1}{0}{1}{0}}{}
\ifthenelse{\equal{#1}{417039}}{\MHB{0}{0}{0}{0}{1}{0}{0}}{}
\ifthenelse{\equal{#1}{417043}}{\MHB{1}{1}{0}{0}{0}{1}{1}}{}
\ifthenelse{\equal{#1}{417044}}{\MHB{1}{1}{1}{0}{0}{1}{1}}{}
\ifthenelse{\equal{#1}{417047}}{\MHB{1}{0}{1}{0}{0}{1}{0}}{}
\ifthenelse{\equal{#1}{41704703x2}}{\MHB{1}{0}{1}{0}{0}{1}{0}\Vertex(11.34,25){\dotsz}}{}
\ifthenelse{\equal{#1}{41704707x2}}{\MHB{1}{0}{1}{0}{0}{1}{0}\Vertex(20,40){\dotsz}}{}
\ifthenelse{\equal{#1}{417048}}{\MHB{1}{1}{0}{0}{1}{0}{0}}{}
\ifthenelse{\equal{#1}{41704804x2}}{\MHB{1}{1}{0}{0}{1}{0}{0}\Vertex(20,10){\dotsz}}{}
\ifthenelse{\equal{#1}{41704806x2}}{\MHB{1}{1}{0}{0}{1}{0}{0}\Vertex(20,30){\dotsz}}{}
\ifthenelse{\equal{#1}{417049}}{\MHB{1}{1}{1}{0}{1}{0}{0}}{}
\ifthenelse{\equal{#1}{417051}}{\MHB{1}{0}{1}{1}{1}{0}{0}}{}
\ifthenelse{\equal{#1}{417052}}{\MHB{1}{0}{1}{1}{0}{0}{1}}{}
\ifthenelse{\equal{#1}{41807}}{\MIA{1}{0}{0}{0}{0}{1}{1}{0}}{}
\ifthenelse{\equal{#1}{418033}}{\MIB{1}{0}{0}{0}{1}{0}{0}{0}}{}
\ifthenelse{\equal{#1}{418034}}{\MIB{0}{1}{0}{1}{0}{0}{0}{0}}{}
\ifthenelse{\equal{#1}{418036}}{\MIB{0}{0}{1}{1}{0}{0}{0}{1}}{}
\ifthenelse{\equal{#1}{418037}}{\MIB{1}{0}{0}{0}{1}{1}{0}{0}}{}
\ifthenelse{\equal{#1}{418040}}{\MIB{1}{0}{1}{0}{0}{1}{0}{1}}{}
\ifthenelse{\equal{#1}{418042}}{\MIB{0}{1}{1}{0}{0}{1}{0}{1}}{}
\ifthenelse{\equal{#1}{418044}}{\MIB{1}{0}{1}{0}{0}{0}{1}{1}}{}
\ifthenelse{\equal{#1}{418048}}{\MIB{1}{1}{1}{0}{0}{0}{0}{1}}{}
\ifthenelse{\equal{#1}{41908}}{\MJC{0}{0}{0}{1}{0}{0}{0}{0}{0}}{}
\ifthenelse{\equal{#1}{419010}}{\MJC{1}{0}{0}{0}{1}{1}{0}{0}{0}}{}
\ifthenelse{\equal{#1}{419011}}{\MJC{1}{0}{0}{0}{0}{0}{1}{1}{1}}{}
\ifthenelse{\equal{#1}{419012}}{\MJC{1}{0}{0}{0}{1}{0}{1}{1}{1}}{}
\ifthenelse{\equal{#1}{41901205x2}}{\MJC{1}{0}{0}{0}{1}{0}{1}{1}{1}\Vertex(40,20){\dotsz}}{}
\ifthenelse{\equal{#1}{41901209x2}}{\MJC{1}{0}{1}{0}{1}{0}{0}{1}{1}\Vertex(0,20){\dotsz}}{}
\ifthenelse{\equal{#1}{419017}}{\MJB{1}{1}{0}{1}{1}{0}{1}{0}{0}}{}
\ifthenelse{\equal{#1}{41808}}{\MIA{1}{0}{0}{1}{0}{1}{0}{1}}{}
\ifthenelse{\equal{#1}{418045}}{\MIB{1}{1}{1}{1}{0}{0}{1}{0}}{}
\ifthenelse{\equal{#1}{418046}}{\MIB{1}{1}{1}{1}{0}{0}{0}{1}}{}
\ifthenelse{\equal{#1}{418047}}{\MIB{1}{0}{1}{1}{0}{0}{0}{1}}{}
\ifthenelse{\equal{#1}{41905}}{\MJB{1}{0}{0}{0}{1}{0}{0}{0}{1}}{}
\ifthenelse{\equal{#1}{419019}}{\MJB{1}{0}{0}{1}{1}{0}{1}{0}{1}}{}
\ifthenelse{\equal{#1}{41904}}{\MJB{0}{0}{0}{1}{1}{0}{0}{0}{1}}{}
\ifthenelse{\equal{#1}{41904sp}}{\MJB{0}{0}{0}{1}{1}{0}{0}{0}{1}\ArrowLine(3,15)(3,25)\Text(8,20)[]{\momsz$p$}\ArrowLine(23.67,15)(23.67,5)\Text(18.67,10)[]{\momsz$q$}}{}
\ifthenelse{\equal{#1}{41906}}{\MJB{1}{0}{0}{1}{1}{0}{0}{0}{1}}{}
\ifthenelse{\equal{#1}{41906sp}}{\MJB{1}{0}{0}{1}{1}{0}{0}{0}{1}\ArrowLine(3,15)(3,25)\Text(8,20)[]{\momsz$p$}\ArrowLine(23.67,15)(23.67,5)\Text(18.67,10)[]{\momsz$q$}}{}
\ifthenelse{\equal{#1}{41907}}{\MJB{1}{0}{0}{1}{0}{0}{0}{1}{1}}{}
\ifthenelse{\equal{#1}{41907sp}}{\MJB{1}{0}{0}{1}{0}{0}{0}{1}{1}\ArrowLine(3,15)(3,25)\Text(8,20)[]{\momsz$p$}\ArrowLine(23.67,15)(23.67,5)\Text(18.67,10)[]{\momsz$q$}}{}
\ifthenelse{\equal{#1}{419020}}{\MJB{0}{1}{0}{1}{1}{0}{0}{1}{0}}{}
\ifthenelse{\equal{#1}{419020sp}}{\MJB{0}{1}{0}{1}{1}{0}{0}{1}{0}\ArrowLine(3,15)(3,25)\Text(8,20)[]{\momsz$p$}\ArrowLine(23.67,35)(23.67,25)\Text(18.67,30)[]{\momsz$q$}}{}
\ifthenelse{\equal{#1}{419021sp}}{\MJB{1}{1}{0}{1}{1}{0}{0}{1}{0}\ArrowLine(3,15)(3,25)\Text(8,20)[]{\momsz$p$}\ArrowLine(23.67,35)(23.67,25)\Text(18.67,30)[]{\momsz$q$}}{}
\ifthenelse{\equal{#1}{419010sp}}{\MJC{1}{0}{0}{0}{1}{1}{0}{0}{0}\ArrowLine(5.5,14.1)(14.17,19.1)\Text(14.5,29.5)[]{\momsz$p$}\ArrowLine(7,32.5)(15.67,37.5)\Text(13,10)[]{\momsz$q$}}{}
\ifthenelse{\equal{#1}{419011sp}}{\MJC{1}{0}{0}{0}{0}{0}{1}{1}{1}\ArrowLine(5.5,14.1)(14.17,19.1)\Text(14.5,29.5)[]{\momsz$p$}\ArrowLine(7,32.5)(15.67,37.5)\Text(13,10)[]{\momsz$q$}}{}
\ifthenelse{\equal{#1}{419012sp}}{\MJC{1}{0}{0}{0}{1}{0}{1}{1}{1}\ArrowLine(5.5,14.1)(14.17,19.1)\Text(14.5,29.5)[]{\momsz$p$}\ArrowLine(7,32.5)(15.67,37.5)\Text(13,10)[]{\momsz$q$}}{}
\ifthenelse{\equal{#1}{Tf52}}{\MIA{0}{1}{1}{1}{1}{0}{1}{1}}{}
\ifthenelse{\equal{#1}{Tf54}}{\MJC{1}{0}{1}{0}{1}{0}{1}{1}{1}\ArrowLine(5.5,14.1)(14.17,19.1)\Text(14.5,29.5)[]{\momsz$p$}\ArrowLine(7,32.5)(15.67,37.5)\Text(13,10)[]{\momsz$q$}}{}
\ifthenelse{\equal{#1}{Tf61}}{\MIB{1}{1}{1}{1}{0}{0}{1}{1}}{}
\ifthenelse{\equal{#1}{Tf62}}{\MJB{0}{1}{1}{1}{0}{1}{0}{1}{1}}{}
\ifthenelse{\equal{#1}{Tf64}}{\MIB{0}{1}{0}{1}{0}{1}{0}{1}}{}
\ifthenelse{\equal{#1}{Tf71}}{\MIB{0}{0}{0}{1}{0}{0}{1}{1}}{}
\ifthenelse{\equal{#1}{Tf72}}{\MIB{1}{1}{0}{0}{1}{0}{1}{1}}{}
\ifthenelse{\equal{#1}{40401}}{\Vertex(20,20){\vtxsz}\Line(20,20)(32.35,25.11)\Line(20,20)(32.35,14.89)\CArc(34.46,20)(5.54,-112.5,112.5)\Line(20,20)(7.65,25.11)\Line(20,20)(7.65,14.89)\CArc(5.54,20)(5.54,67.5,-67.5)\Line(20,20)(25.11,32.35)\Line(20,20)(14.89,32.35)\CArc(20,34.46)(5.54,-22.5,202.5)\Line(20,20)(25.11,7.65)\Line(20,20)(14.89,7.65)\CArc(20,5.54)(5.54,157.5,22.5)}{}
\ifthenelse{\equal{#1}{40523}}{\Vertex(20,0){\vtxsz}\Vertex(20,30){\vtxsz}\CArc(20,15)(15,-90,90)\CArc(20,15)(15,90,270)\CArc(20,35)(5,0,360)\DashCArc(40,15)(25,143.13,216.87){\dash}\DashCArc(0,15)(25,-36.87,36.87){\dash}}{}
\ifthenelse{\equal{#1}{40524}}{\Vertex(20,0){\vtxsz}\Vertex(20,30){\vtxsz}\CArc(20,15)(15,-90,90)\CArc(20,15)(15,90,270)\CArc(20,35)(5,0,360)\CArc(40,15)(25,143.13,216.87)\CArc(0,15)(25,-36.87,36.87)}{}
\ifthenelse{\equal{#1}{40504}}{\MFA{1}{0}{0}{0}{1}}{}
\ifthenelse{\equal{#1}{40505}}{\MFA{1}{1}{0}{1}{1}}{}
\ifthenelse{\equal{#1}{40602}}{\MGB{1}{0}{1}{0}{0}{1}}{}
\ifthenelse{\equal{#1}{40618}}{\MGB{1}{1}{0}{0}{0}{0}}{}
\ifthenelse{\equal{#1}{406111}}{\MGB{1}{1}{1}{1}{1}{1}}{}
\ifthenelse{\equal{#1}{406112}}{\MGB{0}{0}{1}{1}{1}{1}}{}
\ifthenelse{\equal{#1}{407117}}{\Vertex(5.86,5.86){\vtxsz}\Vertex(34.14,5.86){\vtxsz}\Vertex(5.86,34.14){\vtxsz}\Vertex(34.14,34.14){\vtxsz}\CArc(20,20)(20,0,360)\DashLine(5.86,5.86)(34.14,34.14){\dash}\DashLine(34.14,5.86)(5.86,34.14){\dash}\DashLine(5.86,34.14)(34.14,34.14){\dash}}{}
\ifthenelse{\equal{#1}{407118}}{\Vertex(20,0){\vtxsz}\Vertex(13.33,20){\vtxsz}\Vertex(26.67,20){\vtxsz}\Vertex(20,40){\vtxsz}\CArc(20,20)(20,0,360)\CArc(46.66,20)(33.33,143.13,180)\CArc(-6.66,20)(33.33,0,36.87)\DashCArc(46.66,20)(33.33,180,216.87){\dash}\DashCArc(-6.66,20)(33.33,-36.87,0){\dash}\Line(13.33,20)(26.67,20)}{}
\ifthenelse{\equal{#1}{40802}}{\MIB{1}{0}{0}{0}{1}{1}{0}{1}}{}
\ifthenelse{\equal{#1}{40903}}{\MJC{1}{0}{1}{0}{1}{0}{1}{1}{1}}{}
}

\newcommand{\TN}[1]{
\ifthenelse{\equal{#1}{41501}}{T_{5,5}}{}
\ifthenelse{\equal{#1}{41502}}{T_{5,6}}{}
\ifthenelse{\equal{#1}{41503}}{T_{5,7}}{}
\ifthenelse{\equal{#1}{41504}}{T_{5,8}}{}
\ifthenelse{\equal{#1}{4150403x3}}{T_{5,9}}{}
\ifthenelse{\equal{#1}{41505}}{T_{5,10}}{}
\ifthenelse{\equal{#1}{4150401x2}}{T_{5,11}}{}
\ifthenelse{\equal{#1}{416010}}{T_{6,5}}{}
\ifthenelse{\equal{#1}{416018}}{T_{6,6}}{}
\ifthenelse{\equal{#1}{41604}}{T_{6,7}}{}
\ifthenelse{\equal{#1}{41605}}{T_{6,8}}{}
\ifthenelse{\equal{#1}{416023}}{T_{6,9}}{}
\ifthenelse{\equal{#1}{41602303x2}}{T_{6,10}}{}
\ifthenelse{\equal{#1}{416025}}{T_{6,11}}{}
\ifthenelse{\equal{#1}{41602503x2}}{T_{6,12}}{}
\ifthenelse{\equal{#1}{416024}}{T_{6,13}}{}
\ifthenelse{\equal{#1}{41602403x2}}{T_{6,14}}{}
\ifthenelse{\equal{#1}{416011}}{T_{6,15}}{}
\ifthenelse{\equal{#1}{416017}}{T_{6,16}}{}
\ifthenelse{\equal{#1}{416022}}{T_{6,17}}{}
\ifthenelse{\equal{#1}{416021}}{T_{6,18}}{}
\ifthenelse{\equal{#1}{41603}}{T_{6,19}}{}
\ifthenelse{\equal{#1}{417019}}{T_{7,3}}{}
\ifthenelse{\equal{#1}{417039}}{T_{7,4}}{}
\ifthenelse{\equal{#1}{417044}}{T_{7,5}}{}
\ifthenelse{\equal{#1}{417043}}{T_{7,6}}{}
\ifthenelse{\equal{#1}{417049}}{T_{7,7}}{}
\ifthenelse{\equal{#1}{417051}}{T_{7,8}}{}
\ifthenelse{\equal{#1}{417052}}{T_{7,9}}{}
\ifthenelse{\equal{#1}{417047}}{T_{7,10}}{}
\ifthenelse{\equal{#1}{41704707x2}}{T_{7,11}}{}
\ifthenelse{\equal{#1}{417048}}{T_{7,12}}{}
\ifthenelse{\equal{#1}{41704804x2}}{T_{7,13}}{}
\ifthenelse{\equal{#1}{41704806x2}}{---}{}
\ifthenelse{\equal{#1}{417030}}{T_{7,14}}{}
\ifthenelse{\equal{#1}{417036}}{T_{7,15}}{}
\ifthenelse{\equal{#1}{417038}}{T_{7,16}}{}
\ifthenelse{\equal{#1}{41704703x2}}{T_{7,17}}{}
\ifthenelse{\equal{#1}{418040}}{T_{8,2}}{}
\ifthenelse{\equal{#1}{418042}}{T_{8,3}}{}
\ifthenelse{\equal{#1}{418044}}{T_{8,4}}{}
\ifthenelse{\equal{#1}{418048}}{T_{8,5}}{}
\ifthenelse{\equal{#1}{418036}}{T_{8,6}}{}
\ifthenelse{\equal{#1}{418037}}{T_{8,7}}{}
\ifthenelse{\equal{#1}{418033}}{T_{8,8}}{}
\ifthenelse{\equal{#1}{418034}}{T_{8,9}}{}
\ifthenelse{\equal{#1}{41807}}{T_{8,10}}{}
\ifthenelse{\equal{#1}{41908}}{T_{9,2}}{}
\ifthenelse{\equal{#1}{419012}}{T_{9,3}}{}
\ifthenelse{\equal{#1}{41901205x2}}{T_{9,4}}{}
\ifthenelse{\equal{#1}{41901209x2}}{---}{}
\ifthenelse{\equal{#1}{419011}}{T_{9,5}}{}
\ifthenelse{\equal{#1}{419010}}{T_{9,6}}{}
\ifthenelse{\equal{#1}{419017}}{T_{9,7}}{}
\ifthenelse{\equal{#1}{41808}}{T_{5,8}^f}{}
\ifthenelse{\equal{#1}{418045}}{T_{6,14}^f}{}
\ifthenelse{\equal{#1}{418046}}{T_{7,15}^f}{}
\ifthenelse{\equal{#1}{418047}}{T_{7,6}^f}{}
\ifthenelse{\equal{#1}{41905}}{T_{6,15}^f}{}
\ifthenelse{\equal{#1}{419019}}{T_{6,9}^f}{}
\ifthenelse{\equal{#1}{41904}}{T_{6,16}^f}{}
\ifthenelse{\equal{#1}{41904sp}}{T_{6,19}^f}{}
\ifthenelse{\equal{#1}{41906}}{T_{6,13}^f}{}
\ifthenelse{\equal{#1}{41906sp}}{T_{6,18}^f}{}
\ifthenelse{\equal{#1}{41907}}{T_{7,11}^f}{}
\ifthenelse{\equal{#1}{41907sp}}{T_{7,14}^f}{}
\ifthenelse{\equal{#1}{419020}}{T_{5,9}^f}{}
\ifthenelse{\equal{#1}{419020sp}}{T_{5,10}^f}{}
\ifthenelse{\equal{#1}{419021sp}}{T_{6,12}^f}{}
\ifthenelse{\equal{#1}{419010sp}}{T_{6,17}^f}{}
\ifthenelse{\equal{#1}{419011sp}}{T_{6,10}^f}{}
\ifthenelse{\equal{#1}{419012sp}}{T_{9,4}^f}{}
\ifthenelse{\equal{#1}{Tf52}}{T_{5,2}^f}{}
\ifthenelse{\equal{#1}{Tf54}}{T_{5,4}^f}{}
\ifthenelse{\equal{#1}{Tf61}}{T_{6,1}^f}{}
\ifthenelse{\equal{#1}{Tf62}}{T_{6,2}^f}{}
\ifthenelse{\equal{#1}{Tf64}}{T_{6,4}^f}{}
\ifthenelse{\equal{#1}{Tf71}}{T_{7,1}^f}{}
\ifthenelse{\equal{#1}{Tf72}}{T_{7,2}^f}{}
\ifthenelse{\equal{#1}{40401}}{T_{4,1}}{}
\ifthenelse{\equal{#1}{40523}}{T_{5,1}}{}
\ifthenelse{\equal{#1}{40524}}{T_{5,2}}{}
\ifthenelse{\equal{#1}{40504}}{T_{5,3}}{}
\ifthenelse{\equal{#1}{40505}}{T_{5,4}}{}
\ifthenelse{\equal{#1}{40602}}{T_{6,4}}{}
\ifthenelse{\equal{#1}{40618}}{T_{6,3}}{}
\ifthenelse{\equal{#1}{406111}}{T_{6,1}}{}
\ifthenelse{\equal{#1}{406112}}{T_{6,2}}{}
\ifthenelse{\equal{#1}{407117}}{T_{7,2}}{}
\ifthenelse{\equal{#1}{407118}}{T_{7,1}}{}
\ifthenelse{\equal{#1}{40802}}{T_{8,1}}{}
\ifthenelse{\equal{#1}{40903}}{T_{9,1}}{}
}

\newcommand{\TopL}[1]{\raisebox{-0.5mm}[10mm][0mm]{\begin{minipage}[t][0mm][c]{17mm}\begin{center}$\underset{\parbox[b][0mm][c]{15mm}{\begin{center}\scriptsize$\TN{#1}$\end{center}}}{\begin{picture}(\xdim,\ydim)(\xoff,\yoff)\Topo{#1}\end{picture}}$\end{center}\end{minipage}}}
\newcommand{\TopF}[1]{\raisebox{-0.5mm}[10mm][10mm]{\begin{minipage}[t][0mm][c]{17mm}\begin{center}$\underset{\parbox[b][0mm][c]{15mm}{\begin{center}\scriptsize$\TN{#1}$\end{center}}}{\begin{picture}(\xdim,\ydim)(\xoff,\yoff)\Topo{#1}\end{picture}}$\end{center}\end{minipage}}}

\newcommand{\scalarp}{p \cdot q \times\hspace{-1mm}}
\newcommand{\Log}[2]{\log^{#2}(#1)}
\newcommand{\ep}{\epsilon}
\newcommand{\Exp}[1]{e^{#1}}
\newcommand{\Egamma}{\gamma_{E}}
\newcommand{\PGamma}[2]{\Gamma^{#2}\!\left(#1\right)}

\newcommand{\z}[1]{\zeta_{#1}}
\newcommand{\pa}{\text{Li}_4\left(\tfrac{1}{2}\right)}
\newcommand{\pb}[1]{s_{#1}}
\newcommand{\pc}{\text{Li}_5\left(\tfrac{1}{2}\right)}
\renewcommand{\*}{\,}
\def\rela{.5ex} % line pitch within a relation in appendix B
\def\relb{2ex}  % line pitch between two relations in appendix B

\begin{document}

\begin{titlepage}

{
\centerline{\normalsize\hfill  SFB/CPP-06-50}
\centerline{\normalsize\hfill       TTP06-30}
\centerline{\normalsize\hfill hep-ph/0611244}
\centerline{\normalsize\hfill November 2006}
\baselineskip 11pt
{}
}

\vspace{0.5cm}
\begin{center}
  \begin{Large}
  \begin{bf}
Standard and $\boldsymbol{\ep}$-finite Master Integrals\\
for the $\boldsymbol{\rho}$-Parameter
  \end{bf}
  \end{Large}

  \vspace{0.8cm}

  \begin{large}
      M.~Faisst$^a$, P. Maierh\"ofer$^a$ and C.~Sturm$^b$
  \end{large}
  \vskip .7cm

{\small $^a$ {\em Institut f\"ur Theoretische Teilchenphysik,
  Universit\"at Karlsruhe, D-76128 Karlsruhe, Germany}}

{\small $^b$ {\em  Dipartimento di Fisica Teorica, Universit{\`a} di
    Torino, Italy\\   INFN, Sezione di Torino, Italy}}

        \vspace{0.8cm}
{\bf Abstract}
\end{center}
\begin{quotation}
\noindent
We have constructed an $\ep$-finite basis of master integrals for all
new types of one-scale tadpoles which appear in the calculation of the
four-loop QCD corrections to the electroweak $\rho$-parameter. 
Using transformation rules from the $\ep$-finite basis to the standard
``minimal-number-of-lines'' basis, we obtain as a by-product analytical
expressions for few leading terms of the $\ep$-expansion of all members
of the standard basis. The new master integrals have been computed with
the help of the Pad\'e method and by use of difference equations
independently.\\[0.8cm]
\end{quotation}
\end{titlepage}

%%%%%%%%%%%%%%%%%%%%%%%%%%%%%%%%%%%%%%%%%%%%%%%%%%%%%%%%%%%%%%%%%%%%%%%%%%%%%%%%
%%%%%%%%%%%%%%%%%%%%%%%%%%%%%%%%%%%%%%%%%%%%%%%%%%%%%%%%%%%%%%%%%%%%%%%%%%%%%%%%
%%%%%%%%%%%%%%%%%%%%%%%%%%%%%%%%%%%%%%%%%%%%%%%%%%%%%%%%%%%%%%%%%%%%%%%%%%%%%%%%
%%%%%%%%%%%%%%%%%%%%%%%%%%%%%%%%%%%%%%%%%%%%%%%%%%%%%%%%%%%%%%%%%%%%%%%%%%%%%%%%
%%%%%%%%%%%%%%%%%%%%%%%%%%%%%%%%%%%%%%%%%%%%%%%%%%%%%%%%%%%%%%%%%%%%%%%%%%%%%%%%
\section{Introduction\label{sec:Intro}}

\begin{figure}[!ht]
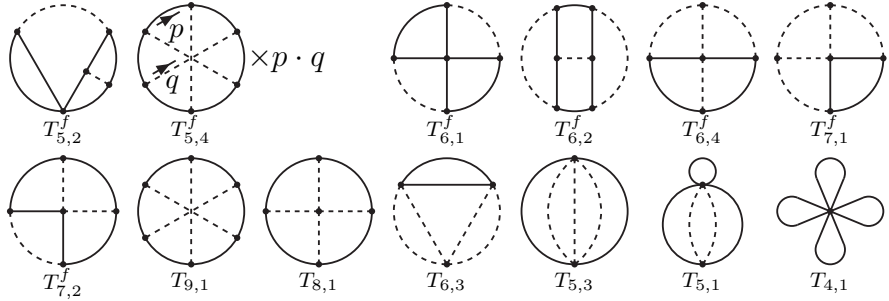

\begin{center}
$\TopF{Tf52}\TopF{Tf54}\parbox[t][0mm][t]{17mm}{$\hspace*{-1mm}\times
p\cdot{}q$}\TopF{Tf61}\TopF{Tf62}\TopF{Tf64}\TopF{Tf71}$\\
$\TopF{Tf72}\TopF{40903}\TopF{40802}\TopF{40618}\TopF{40504}\TopF{40523}\TopF{40401}$\\
\end{center}
%%%%%%%%%%%%%%%%%%%%%%%%%%%%%%%%%%%%%%%%%%%%%%%%%%%%%%%%%%%%%%%%%%%%%%%
  \caption{\label{fig:epfb}Diagrams of the subset of 13 master integrals
    of the $\ep$-finite basis addressed in ref.~\cite{Chetyrkin:2006dh}.
    Note that the last four diagrams are known completely analytically
    and are not considered in the construction of the $\ep$-finite
    basis. 
}
\end{figure}

In many multi-loop calculations the standard method to calculate
physical observables is to use the traditional
integration-by-parts~(IBP) method in combination
with Laporta's algorithm~\cite{Laporta:2001dd,Laporta:1996mq} in order
to reduce all appearing integrals to a small set of master integrals.
Once this reduction is completed one is left with the calculation of the
master integrals, addressed in this work. \\
One physical quantity which has recently been evaluated applying these
methods is the contribution from  top- and bottom-quarks
to the $\rho$-parameter at four-loop order in perturbative
QCD~\cite{Chetyrkin:2006bj}. A subsequent independent evaluation of
the same quantity has been performed in ref.~\cite{Boughezal:2006xk}.
This completes the partial result from the so-called singlet term already
determined in ref.~\cite{Schroder:2005db}.
For  the $\rho$-parameter,
the result can be expressed in terms of 63 four-loop tadpole master integrals.
A subset of 13 master integrals was already required in earlier four-loop
calculations, like the determination of the matching condition for the
strong coupling constant $\alpha_s$ at a heavy quark threshold in the
modified minimal subtraction
scheme~\cite{Chetyrkin:2005ia,Schroder:2005hy} and the evaluation of the two
lowest terms of the Taylor expansion of the vacuum polarization function
\cite{Chetyrkin:2006xg,Boughezal:2006px}.  
This subset of master integrals has been evaluated in ref.~\cite{Schroder:2005va}
 with high precision
using the method of difference equations
\cite{Laporta:1996mq,Laporta:2000dc,Laporta:2001dd,Laporta:2001rc,Laporta:2002pg}.
 {\em All}  results relevant for the four-loop
calculation of ref.~\cite{Schroder:2005va} have  been
confirmed with a completely different method (see below) in
ref.~\cite{Chetyrkin:2006dh}. Some of these master integrals have also
been found in 
refs.~\cite{Broadhurst:1992fi,Broadhurst:1996az,Laporta:2002pg,Chetyrkin:2004fq,Kniehl:2005yc,Schroder:2005db,Bejdakic:2006vg,Kniehl:2006bf,Kniehl:2006bg}. \\
A different method, based on the idea of the $\ep$-finite basis, has been suggested in
ref.~\cite{Chetyrkin:2006dh}. In this approach one avoids
so-called spurious poles in the coefficient functions, which multiply the
master integrals. These poles may arise in general while solving the linear system
of IBP equations, when a division by $\ep=(4-d)/2$ occurs. Master integrals
which have a spurious pole as coefficient need to be evaluated deeper in
the $\ep$-expansion. The calculation of each
additional order in this expansion of a master integral is in general
increasingly tedious. In contrast, the approach of the $\ep$-finite basis exploits
the freedom in the choice of master integrals in order to select a basis of
master integrals in such a way that the coefficients, being functions in the
space time dimension $d$, are finite in the limit $\ep\rightarrow 0$.  In
ref.~\cite{Chetyrkin:2006dh} the $\ep$-finite basis, shown in
fig.~\ref{fig:epfb}, has been constructed for the subset of 13 master
integrals discussed above.
%%%%%%%%%

All of them have been calculated semi-analytically. For the problem of the
four-loop QCD corrections to the $\rho$-parameter an $\ep$-finite basis with
more elements  is required which is constructed in the following.
The corresponding master integrals are particularly suited for the
evaluation by a semi-numerical method based on Pad\'e
approximations \cite{Faisst:2004kz,Faisst:diss,Chetyrkin:2006dh}.
Most of the results are checked independently by the method of
difference equations and have been used for the evaluation of the 
$\rho$-parameter in ref.~\cite{Chetyrkin:2006bj}.
They are in full agreement with those recently published in \cite{Boughezal:2006xk}.

This paper is structured as follows. In the next section we present the
results for the master integrals in the $\epsilon$-finite basis,
calculated by the Pad\'e method. Then, in section~\ref{sec:standard}, we
give results for the master integrals in the standard basis
evaluated by difference equations. In this
case we also include the analytical information obtained in 
section~\ref{sec:epf}. Our summary and conclusions are given in
section~\ref{sec:conc}.

\section{The master integrals in the $\boldsymbol{\ep}$-finite basis\label{sec:epf}}

For the construction of the $\ep$-finite basis we follow the algorithm
described in reference \cite{Chetyrkin:2006dh}. 
In the evaluation of the $\rho$-parameter we have again a vast choice in defining the $\ep$-finite
basis with elements suitable for the Pad\'e method.
We exclude 14 diagrams from  the construction of the $\ep$-finite basis,
which are known to high orders in~$\ep$ analytically or even completely
analytically. Four of them are depicted in
fig.~\ref{fig:epfb} ($T_{4,1},T_{5,1},T_{5,3},T_{6,3}$), the remaining ones
are listed in appendix~\ref{app:analytic}.

The application of the method
\cite{Faisst:2004kz,Faisst:diss,Chetyrkin:2006dh} to the actual
calculation of the $\ep$-finite integrals is straightforward. The only
adjustment necessary arises from the appearance of massless cuts
inherent in many of the relevant diagrams. In addition to the function
used to subtract the high energy logarithms, a second function was
introduced to subtract the logarithms appearing in the calculation of
diagrams with a massless cut in the low energy limit. This approach was
also verified by recalculating the diagrams presented in
fig.~\ref{fig:epfb} (e.g. $T_{6,4}^f$); this time performing the cut in
such a way that a self energy diagram with a massless cut arises. Full
agreement between the two approaches was observed.  Following this
procedure we again calculate all divergent parts of the diagrams
analytically, and the constant part in the limit $\ep \to 0$
numerically.  All the digits of our numerical results in
eqs.~(\ref{41808})-(\ref{419010sp}) are valid digits. The analytical
information of the expansion then allows to study the cancellation of
all divergences in the physical problem during the renormalization
procedure analytically. Just as in the problem of the diagonal current,
some diagrams are simultaneously members of the standard and the
$\ep$-finite basis. These integrals are discussed in
section~\ref{sec:standard}. Now we give the results for the integrals
which are different in the two bases, i.e. the replaced integrals due to
the existence of a spurious pole. As in ref.~\cite{Chetyrkin:2006dh} the
name of the $\ep$-finite master integral is deduced by using the name of
the original master integral, which has been replaced, and the
additional superscript letter ``$f$''.
The loop integrals are defined as follows,
\begin{equation}
J = \int [dk_1] [dk_2] [dk_3] [dk_4] 
\frac{1}{D_1 D_2 \dots D_{N_d}} \,, 
\end{equation}
with $N_d$ denominators $D_j=(p_j^2-m_j^2)$. The momenta $p_j$ are
defined by the diagrams shown below as a linear combination of the four loop 
momenta $k_i$. For briefness we set the single mass scale $m$ of the
integrals in the following $m=1$, thus $m_j\in\{0,1\}$. Dashed(solid) lines in
all the diagrams denote massless(massive) propagators. The integration
measure is given by $[dk_j]=e^{\ep\*\Egamma}/(i\*\pi^{d/2})\*d^dk_j$, where
$\Egamma=0.577216\dots$ is the Euler's constant. In this conventions the
one-loop tadpole reads: 
\begin{equation}
T_{1}=-e^{\ep\*\Egamma}\*\Gamma(1-d/2)=
  \frac{1}{\ep} 
+ 1 
+ \ep\*\left(1 + \frac{\pi^2}{12}\right)  
+ \ep^2\*\left(1 + \frac{\pi^2}{12} - \frac{\z3}{3}\right)
+ \mathcal{O}(\ep^3)\,.
\end{equation}
The final expressions for the remaining master integrals in the $\ep$-finite basis read:
\begin{align}
\TopL{41808} 
&= 
  \frac{3\*\z3}{2\*\ep^2} 
- \frac{1}{\ep}\*\Big(
  \frac{91}{360}\*\pi^4
- \frac{15}{2}\*\z3
- \frac{4}{3}\*\Log{2}{4}
\nonumber \\ \displaybreak[1] 
&\phantom{=} 
+ \frac{4}{3}\*\pi^2\*\Log{2}{2}
- 32\*\pa\Big)
+ 43.4562683
+ \mathcal{O}(\ep)
\label{41808}\,,\displaybreak[1]\\%%%%%%%%%%%%%%%%%%%%%%%%%%%%%%%%%%%%%%%%%%%%%%%%
%%%%%%%%%%%%%%%%%%%%%%%%%%%%%%%%%%%%%%%%%%%%%%%%%%%%%%%%%%%%%%%%%%%%%
\TopF{418045} 
&= 
  \frac{5\*\z5}{\ep}
- 23.2862341
+ \mathcal{O}(\ep)
\,,\displaybreak[1]\\%%%%%%%%%%%%%%%%%%%%%%%%%%%%%%%%%%%%%%%%%%%%%%%%
\TopF{418046} 
&= 
  \frac{5\*\z5}{\ep}
- 24.8172810
+ \mathcal{O}(\ep)
\,,\displaybreak[1]\\%%%%%%%%%%%%%%%%%%%%%%%%%%%%%%%%%%%%%%%%%%%%%%%%
\TopF{418047} 
&= 
  \frac{5\*\z5}{\ep}
- 18.9284641
+ \mathcal{O}(\ep)
\,,\displaybreak[1]\\%%%%%%%%%%%%%%%%%%%%%%%%%%%%%%%%%%%%%%%%%%%%%%%%
%%%%%%%%%%%%%%%%%%%%%%%%%%%%%%%%%%%%%%%%%%%%%%%%%%%%%%%%%%%%%%%%%%%%%
\TopF{419019} 
&= 
- 3.2095107
+ \mathcal{O}(\ep)
\,,\displaybreak[1]\\%%%%%%%%%%%%%%%%%%%%%%%%%%%%%%%%%%%%%%%%%%%%%%%%
\scalarp\TopF{419021sp} 
&= 
  \frac{1}{2\*\ep}\*\left(5\*\z5-3\*\z3\right)
- 3.67204
+ \mathcal{O}(\ep)
\,,\displaybreak[1]\\%%%%%%%%%%%%%%%%%%%%%%%%%%%%%%%%%%%%%%%%%%%%%%%%
\TopF{41906} 
&= 
- 5.1732238
+ \mathcal{O}(\ep)
\,,\displaybreak[1]\\%%%%%%%%%%%%%%%%%%%%%%%%%%%%%%%%%%%%%%%%%%%%%%%%
\scalarp\TopF{41906sp}  
&= 
  \frac{1}{2\*\ep}\*\left(5\*\z5-3\*\z3\right)
- 1.589373
+ \mathcal{O}(\ep)
\,,\displaybreak[1]\\%%%%%%%%%%%%%%%%%%%%%%%%%%%%%%%%%%%%%%%%%%%%%%%%
\TopF{41907} 
&= 
- 5.126697
+ \mathcal{O}(\ep)
\,,\displaybreak[1]\\%%%%%%%%%%%%%%%%%%%%%%%%%%%%%%%%%%%%%%%%%%%%%%%%
\scalarp\TopF{41907sp} 
&= 
  \frac{1}{2\*\ep}\*\left(5\*\z5-3\*\z3\right)
+ 0.0114944
+ \mathcal{O}(\ep)
\,,\displaybreak[1]\\%%%%%%%%%%%%%%%%%%%%%%%%%%%%%%%%%%%%%%%%%%%%%%%%
\TopF{419020} 
&= 
- 6.8481671
+ \mathcal{O}(\ep)
\,,\displaybreak[1]\\%%%%%%%%%%%%%%%%%%%%%%%%%%%%%%%%%%%%%%%%%%%%%%%%
\scalarp\TopF{419020sp} 
&= 
  \frac{1}{2\*\ep}\*\left(5\*\z5-3\*\z3\right)
- 3.272794
+ \mathcal{O}(\ep)
\,,\displaybreak[1]\\%%%%%%%%%%%%%%%%%%%%%%%%%%%%%%%%%%%%%%%%%%%%%%%%
\TopF{41904} 
&=  
- 9.5955369
+ \mathcal{O}(\ep)
\,,\displaybreak[1]\\%%%%%%%%%%%%%%%%%%%%%%%%%%%%%%%%%%%%%%%%%%%%%%%%
\scalarp\TopF{41904sp}  
&= 
  \frac{1}{2\*\ep}\*\left(5\*\z5-3\*\z3\right)
- 0.80223
+ \mathcal{O}(\ep)
\,,\displaybreak[1]\\%%%%%%%%%%%%%%%%%%%%%%%%%%%%%%%%%%%%%%%%%%%%%%%%
\TopF{41905} 
&= 
- 8.816973
+ \mathcal{O}(\ep)
\,,\displaybreak[1]\\%%%%%%%%%%%%%%%%%%%%%%%%%%%%%%%%%%%%%%%%%%%%%%%%
%%%%%%%%%%%%%%%%%%%%%%%%%%%%%%%%%%%%%%%%%%%%%%%%%%%%%%%%%%%%%%%%%%%%%
\scalarp\TopF{419012sp} 
&= 
- \frac{5\*\z5}{4\*\ep}
+ 6.2801043
+ \mathcal{O}(\ep)
\,,\displaybreak[1]\\%%%%%%%%%%%%%%%%%%%%%%%%%%%%%%%%%%%%%%%%%%%%%%%%
\scalarp\TopF{419011sp} 
&=  
- \frac{5\*\z5}{4\*\ep} 
+ 5.68333946
+ \mathcal{O}(\ep)
\,,\displaybreak[1]\\%%%%%%%%%%%%%%%%%%%%%%%%%%%%%%%%%%%%%%%%%%%%%%%%
\scalarp\TopF{419010sp} 
&= 
- \frac{5\*\z5}{4\*\ep} 
- 4.872849
+ \mathcal{O}(\ep)
\label{419010sp}\,,%\displaybreak[1]\\%%%%%%%%%%%%%%%%%%%%%%%%%%%%%%%%%%%%%%%%%%%%%%%%
\end{align}
%%%%%
with the Riemann zeta function $\zeta_n$ and the poly-logarithm function
$\mbox{Li}_n(z)$ being defined by: 
\begin{equation}
\zeta_n=\sum_{k=1}^{\infty}\frac{1}{k^n}\quad\mbox{and}\quad
\mbox{Li}_n(z)=\sum_{k=1}^{\infty}\frac{z^k}{k^{n}}\,.
\end{equation}

\section{The master integrals in the standard basis\label{sec:standard}}

The results from the previous section can also be translated into
analytical results for the standard ``minimal-number-of-lines'' basis
for a few leading terms of the $\ep$-expansion by using the
IBP-relations between both bases. Since the order $\ep^0$ of the
$\ep$-finite master integrals in section~\ref{sec:epf} is known
numerically, one can also derive one additional order numerically
through the IBP-relations for the members of the standard basis from the
$\ep$-finite one. On the other hand one can perform a formal expansion
of the standard master integrals:
\begin{equation}
T_{i,j} = \sum_{k=n_{\mathrm{min} }}^{\infty} \ep^k \, T_{i,j}^{(k)}
\label{taylor}
\end{equation}
and insert it into the IBP-relations, which express the $\ep$-finite
master integrals of section~\ref{sec:epf} in terms of the standard
ones. Here we denote each of the master integrals according to the
following rule: after a capital letter ``$T$'' we write the number of
lines (index $i$) in the given diagram. The second number ($j$)
enumerates the different topologies with the same number of lines. If a
particular order $k$ of the $\ep$-expansion of the master integral
$T_{i,j}$ is considered, it is denoted by an additional last index
($k$).  After performing the $\ep$-expansion of the IBP-relations one
compares then the different orders in $\ep$ and obtains thus a linear
system of equations.  Its solution gives additional relations among
particular orders of different master integrals. They are given in
appendix~\ref{app:SpecialRelations}.\\

Another very convenient approach for evaluating master integrals with
eight or less number of lines is given by the numerical solution of
difference equations~\cite{Laporta:1996mq,Laporta:2000dc,%
Laporta:2001dd,Laporta:2001rc,Laporta:2002pg}. This method has been
applied directly to the diagrams of the standard basis providing an
independent check of the results in the previous section and leading to
much higher numerical accuracy. The idea of this method is described in
detail in ref.~\cite{Laporta:2001dd}, we restrict ourself on a brief
sketch. In this approach one of the massive propagators is raised to a
symbolic power $x$
\begin{equation}
J(x) = \int [dk_1] [dk_2] [dk_3] [dk_4] \frac{1}{D_1^x D_2 \dots D_{N_d}} \ ,
\end{equation}
and the IBP-method is used to construct difference equations for the
master integrals by solving a linear system of equations.  The
appearance of two variables, the space-time dimension $d$ and the power
$x$, complicates the solution of the system of IBP-identities. It is solved
with the help of a modified version of the setup described in
ref.~\cite{Sturm:diss} using
{\tt{Form}}\cite{Vermaseren:2000nd,Vermaseren:2002rp,Tentyukov:2006ys}
and {\tt{Fermat}}\cite{Lewis}.
In the case of the $\rho$-parameter also master integrals with 
increased powers of a propagator appear. Then the increased
power is chosen in such a way that it is carried by the propagator
$D_1$, which carries in the above approach the symbolical power~$x$.
In this case the integral is obtained as a by-product of the solution of
the integral without extra power.
The difference equation for $J(x)$ can be written in the form
\begin{equation}
\sum_{j=0}^{R_h} p_j(x) J(x-j) = \sum_{n=1}^{N_k} \sum_{j=0}^{R_i^{(n)}}
q_j^{(n)}(x) K^{(n)}(x-j) \ ,
\end{equation}
where $p_j$ and $q_j^{(n)}$ are polynomials in $x$ and
$d$. $K^{(n)}(x)$ are sub-topologies of $J(x)$, i.~e. topologies with at
least one propagator less, which also satisfy difference equations. At
this point the solutions for the sub-topologies are assumed to be already
known.

The solution is split into several parts determined by using a factorial
series  
\begin{equation}
J_i(x) = \mu_i^x \sum_{s=0}^\infty a_s^{(i)}
\frac{\Gamma(x+1)}{\Gamma(x+1+s-K_i)} 
\end{equation}
as an ansatz. From the difference equation one gets the values
of $\mu_i$ and $K_i$ and a recursion formula for the coefficients
$a_s^{(i)}$. This leads to a particular solution $J_p(x)$ of the
inhomogeneous equation and to one or more homogeneous solutions
$J_h^{(k)}(x)$. The full solution is given by 
\begin{equation}
J(x) = J_p(x) + \sum_k \eta_k(d) J_h^{(k)}(x)
\end{equation}
with functions $\eta_k(d)$, fixed by boundary conditions. The
$\eta_k(d)$ can be determined from the low energy expansion of the
self-energy diagram which one obtains by cutting the line that carries
the power $x$ in the diagram $J(x)$. To obtain a numerical result, the
coefficients $a_s^{(i)}$ are calculated as an expansion in
$\epsilon=(4-d)/2$ up to a pre-defined depth $\epsilon_{max}$. Summing the
series to a specified $s=s_{max}$ and estimating the remainder gives the
numerical result. 

This method provides a possibility to calculate the
$\epsilon$-expansions of master integrals with high numerical precision
up to high orders of $\epsilon$. 

Integrals that we did not calculate by difference equations have been
constructed from the $\ep$-finite basis.  The lowest order, which is
only known numerically, can also be obtained through the IBP-relations
from the $\ep$-finite basis, calculated by the Pad{\'e} approach, but
with less precision. The precision obtained with the Pad{\'e} approach
is marked in the results with an underlined ($\underline{\phantom{0}}$)
digit.  For 11 integrals ($T_{7,10}, T_{7,11}, T_{7,12}, T_{7,13},
T_{8,3}, T_{8,5}, T_{9,3}, T_{9,4}, T_{9,5}, T_{9,6}, T_{9,7}$) more
precise numerical results have in the meantime been obtained in
ref.~\cite{Boughezal:2006xk}. They are appended in the following to our
results and separated by a vertical bar~($|$). All given digits of our
numerical results are valid digits. The results for the master integrals
read:
\begin{align}
\TopL{41504}&= 
  \frac{1}{4\*\ep^4}
+ \frac{1}{\ep^3} 
+ \frac{1}{12\*\ep^2}\*\left(\frac{97}{4} + \pi^2\right)
+ \frac{1}{3\*\ep}\*\left(\frac{833}{96} + \pi^2 - \z3 \right) 
\nonumber\displaybreak[1]\\&
+ \frac{26509}{1728} 
+ \frac{97}{144}\*\pi^2
+ \frac{\pi^4}{12}
- \frac{4}{3}\*\z3
- \frac{11}{2}\*\sqrt{3}\*\pb2
+ 4\*\pb2^2 
\nonumber\displaybreak[1]\\&
+ 80.895\underline{5}061678534024741104898217973159065185099065309\*\ep
\nonumber\displaybreak[1]\\&
+ 1085.28365870727983857702515746382256304748449190947\*\ep^2 
\nonumber\displaybreak[1]\\&
+ 4545.30388413442580391360145657533885328523257544774\*\ep^3 
\nonumber\displaybreak[1]\\&
+ 35998.9938326326556313329606141017307302536969621971\*\ep^4 
\nonumber\displaybreak[1]\\&
+ 134897.307552053604704118477425872393007397735113520\*\ep^5
%\nonumber\displaybreak[1]\\&
+\mathcal{O}(\ep^6)
\label{41504}\,,\displaybreak[1]\\%%%%%%%%%%%%%%%%%%%%%%%%%%%%%%%%%%%%%%%%%%%%%%%%
\TopL{4150403x3}&= 
- \frac{1}{4\*\ep^3}
- \frac{43}{48\*\ep^2} 
- \frac{1}{4\*\ep}\*\left(\frac{51}{8} + \frac{\pi^2}{3}\right)
\nonumber\displaybreak[1]\\&
+ \frac{161}{192}
- \frac{43}{144}\*\pi^2
+ \frac{\z3}{3}
- \frac{3}{2}\*\sqrt{3}\*\pb2
\nonumber\displaybreak[1]\\&
+ 7.3892\underline{4}322701052496152773911803203314613623175263798\*\ep 
\nonumber\displaybreak[1]\\&
+ 101.911510819562293846185747116577265565915688840327\*\ep^2 
\nonumber\displaybreak[1]\\&
+ 413.971474923722132422941214144351086349919086346111\*\ep^3 
\nonumber\displaybreak[1]\\&
+ 2899.99385888238845225724197005268821906921264538028\*\ep^4 
\nonumber\displaybreak[1]\\&
+ 6906.33326400669495615717643151117125986689939048417\*\ep^5
%\nonumber\displaybreak[1]\\&
+\mathcal{O}(\ep^6)
\,,\displaybreak[1]\\%%%%%%%%%%%%%%%%%%%%%%%%%%%%%%%%%%%%%%%%%%%%%%%%
\TopL{41505}&=
  \frac{1}{\ep^4} 
+ \frac{19}{4\*\ep^3}
+ \frac{1}{\ep^2}\*\left(\frac{103}{8} + \frac{\pi^2}{3}\right)
%\nonumber\displaybreak[1]\\&
+ \frac{1}{\ep}\*\left(\frac{341}{16} + \frac{19}{12}\*\pi^2 
     + 3\*\sqrt{3}\*\pb2 
     - \frac{4}{3}\*\z3\right)
\nonumber\displaybreak[1]\\&
+ 56.280301\underline{6}207811654954905101988728760185938023940072
\nonumber\displaybreak[1]\\&
+ 0.62861591651881519024134469355376197437836304764183\*\ep 
\nonumber\displaybreak[1]\\&
- 713.454611137588809603443079328921106975591183151813\*\ep^2 
\nonumber\displaybreak[1]\\&
- 3252.49900904808028994099278455214180709190379330122\*\ep^3 
\nonumber\displaybreak[1]\\&
- 14732.7748604549077668424518963805693923717102434909\*\ep^4 
\nonumber\displaybreak[1]\\&
- 48411.3691880431214632694555990972265902582845398420\*\ep^5
%\nonumber\displaybreak[1]\\&
+\mathcal{O}(\ep^6)
\,,\displaybreak[1]\\%%%%%%%%%%%%%%%%%%%%%%%%%%%%%%%%%%%%%%%%%%%%%%%% 
%%%%%%%%%%%%%%%%%%%%%%%%%%%%%%%%%%%%%%%%%%%%%%%%
%%%%%%%%%%%%%%%%%%%%%%%%%%%%%%%%%%%%%%%%%%%%%%%%
%%%%%%%%%%%%%%%%%%%%%%%%%%%%%%%%%%%%%%%%%%%%%%%%
\TopL{416023}&= 
  \frac{1}{\ep^4} 
+ \frac{25}{4\*\ep^3}
+ \frac{1}{3\*\ep^2}\*\left(\frac{257}{4} + \pi^2\right)
%\nonumber\displaybreak[1]\\&
+ \frac{1}{\ep}\*\left(\frac{433}{12} + \frac{25}{12}\*\pi^2 
   + 6\*\sqrt{3}\*\pb2 
   - \frac{10}{3}\*\z3\right)
\nonumber\displaybreak[1]\\&
+ 13.591650\underline{4}314381310784889154404088589280071858101489
\nonumber\displaybreak[1]\\&
- 595.072765215957129399869660561653897265114979088125\*\ep 
\nonumber\displaybreak[1]\\&
- 6875.84113744124081014720730186766512655050693091046\*\ep^2 
\nonumber\displaybreak[1]\\&
- 31308.1849304592520703889751908067773878059674361777\*\ep^3 
\nonumber\displaybreak[1]\\&
- 192167.397958807277872297783881991122229270292735722\*\ep^4 
\nonumber\displaybreak[1]\\&
- 729545.659023744499039120484965719141090831740292666\*\ep^5
%\nonumber\displaybreak[1]\\&
+\mathcal{O}(\ep^6)
\,,\displaybreak[1]\\%%%%%%%%%%%%%%%%%%%%%%%%%%%%%%%%%%%%%%%%%%%%%%%%
\TopL{41602303x2}&=
  \frac{5}{12\*\ep^4}
+ \frac{7}{4\*\ep^3}
+ \frac{1}{12\*\ep^2}\*\left(43 + \frac{5}{3}\*\pi^2\right)
\nonumber\displaybreak[1]\\&
- \frac{1}{\ep}\*\left(\frac{13}{4}   - \frac{7}{12}\*\pi^2 
     - 2\*\sqrt{3}\*\pb2 + \frac{19}{18}\*\z3\right)
\nonumber\displaybreak[1]\\&
- 37.457464\underline{9}312968260550782554633103343746703536006083
\nonumber\displaybreak[1]\\&
- 296.071923695141317174869536814657610506616897735608\*\ep 
\nonumber\displaybreak[1]\\&
- 2022.20010652903000709521858304176327369282045103655\*\ep^2 
\nonumber\displaybreak[1]\\&
- 8244.01633827202034423181172211035421212582682534110\*\ep^3 
\nonumber\displaybreak[1]\\&
- 46235.4396348606702002905412792127287096490637772429\*\ep^4 
\nonumber\displaybreak[1]\\&
- 162640.942801514076139130843600849619080175249655387\*\ep^5
%\nonumber\displaybreak[1]\\&
+\mathcal{O}(\ep^6)
\,,\displaybreak[1]\\%%%%%%%%%%%%%%%%%%%%%%%%%%%%%%%%%%%%%%%%%%%%%%%%
\TopL{416025}&=
  \frac{7}{12\*\ep^4} 
+ \frac{43}{12\*\ep^3} 
+ \frac{1}{4\*\ep^2}\*\left(47 + \frac{7}{9}\*\pi^2\right)
\nonumber\displaybreak[1]\\&
+ \frac{1}{\ep}\*\left(\frac{187}{12} + \frac{43}{36}\*\pi^2 
   + 6\*\sqrt{3}\*\pb2 + \frac{7}{18}\*\z3\right)
\nonumber\displaybreak[1]\\&
- 51.225\underline{2}212801189152418819912274965922080241128506972
\nonumber\displaybreak[1]\\&
- 477.947460995194619277285562191897453203074025533346\*\ep 
\nonumber\displaybreak[1]\\&
- 6127.93433249128299350151222877004479343657194425540\*\ep^2 
\nonumber\displaybreak[1]\\&
- 23752.7464620411276168084881606893540782888293578810\*\ep^3 
\nonumber\displaybreak[1]\\&
- 161921.400195144895866566378264214720989222786416151\*\ep^4 
\nonumber\displaybreak[1]\\&
- 555837.682068927549027491134693752433283969601726090\*\ep^5
%\nonumber\displaybreak[1]\\&
+\mathcal{O}(\ep^6)
\,,\displaybreak[1]\\%%%%%%%%%%%%%%%%%%%%%%%%%%%%%%%%%%%%%%%%%%%%%%%%
\TopL{41602503x2}&=
  \frac{1}{3\*\ep^4}
+ \frac{4}{3\*\ep^3}
+ \frac{1}{3\*\ep^2}\*\left(7 + \frac{\pi^2}{3}\right)
\nonumber\displaybreak[1]\\&
+ \frac{1}{\ep}\*\left( -\frac{16}{3} + \frac{4}{9}\*\pi^2 
   + 4\*\sqrt{3}\*\pb2 - \frac{4}{9}\*\z3\right)
\nonumber\displaybreak[1]\\&
- 57.585\underline{1}192392905729834630220283359182548958203213894
\nonumber\displaybreak[1]\\&
- 226.906688638637087111211732760488182253117551056418\*\ep 
\nonumber\displaybreak[1]\\&
- 2357.42884823258106273364123808844182706228997262473\*\ep^2 
\nonumber\displaybreak[1]\\&
- 6885.09836097771127634712790473221183461334003969682\*\ep^3 
\nonumber\displaybreak[1]\\&
- 51888.4446361973078707935677603480403504827064470340\*\ep^4 
\nonumber\displaybreak[1]\\&
- 139706.649615179144594349102097042435381107174124316\*\ep^5
%\nonumber\displaybreak[1]\\&
+\mathcal{O}(\ep^6)
\,,\displaybreak[1]\\%%%%%%%%%%%%%%%%%%%%%%%%%%%%%%%%%%%%%%%%%%%%%%%%
\TopL{416024}&=
  \frac{1}{3\*\ep^4}
+ \frac{23}{12\*\ep^3} 
+ \frac{1}{3\*\ep^2}\*\left(\frac{65}{4} + \frac{\pi^2}{3}\right)
\nonumber\displaybreak[1]\\&
+ \frac{1}{\ep}\*\left(\frac{13}{12} + \frac{5}{9}\*\pi^2 
   + 3\*\sqrt{3}\*\pb2 + \frac{20}{9}\*\z3\right)
\nonumber\displaybreak[1]\\&
- 59.198\underline{7}779967727811953138849540094767971219526012487
\nonumber\displaybreak[1]\\&
- 549.042726795028324060762834253615726678116981354786\*\ep 
\nonumber\displaybreak[1]\\&
- 4293.52617296657045021413760744102602226521264919104\*\ep^2 
\nonumber\displaybreak[1]\\&
- 20846.8297417686461476497501335647508755198702169076\*\ep^3 
\nonumber\displaybreak[1]\\&
- 110899.174077030571795400409124084216248915113532175\*\ep^4 
\nonumber\displaybreak[1]\\&
- 470220.656770151179792855460983971995962315522021662\*\ep^5
%\nonumber\displaybreak[1]\\&
+\mathcal{O}(\ep^6)
\,,\displaybreak[1]\\%%%%%%%%%%%%%%%%%%%%%%%%%%%%%%%%%%%%%%%%%%%%%%%%
\TopL{41602403x2}&= 
  \frac{1}{4\*\ep^4}
+ \frac{11}{12\*\ep^3}
+ \frac{1}{12\*\ep^2}\*\left(13 + \pi^2\right)
\nonumber\displaybreak[1]\\&
+ \frac{1}{\ep}\*\left(-\frac{89}{12} + \frac{11}{36}\*\pi^2 
    + 2\*\sqrt{3}\*\pb2 + \frac{17}{6}\*\z3\right)
\nonumber\displaybreak[1]\\&
- 62.926\underline{3}953267520790182858402690009143386475645244229
\nonumber\displaybreak[1]\\&
- 258.692941139129800870656513171840779122656511379367\*\ep 
\nonumber\displaybreak[1]\\&
- 2245.88328900424600912128374855476088078582690149658\*\ep^2 
\nonumber\displaybreak[1]\\&
- 7643.28441760863244533368790482441514380733014950851\*\ep^3 
\nonumber\displaybreak[1]\\&
- 48359.2665176408023251619751028356251676249307751892\*\ep^4 
\nonumber\displaybreak[1]\\&
- 156259.773282065061616264607990167056276913844653462\*\ep^5
%\nonumber\displaybreak[1]\\&
+\mathcal{O}(\ep^6)
\,,\displaybreak[1]\\%%%%%%%%%%%%%%%%%%%%%%%%%%%%%%%%%%%%%%%%%%%%%%%%
\TopL{416011}&=
  \frac{1}{12\*\ep^4}
+ \frac{5}{12\*\ep^3}
+ \frac{1}{12\*\ep^2}\*\left(7 + \frac{\pi^2}{3}\right)
- \frac{1}{6\*\ep}\*\left( \frac{67}{2} 
    + \frac{\pi^2}{6} - \frac{37}{3}\*\z3\right)
\nonumber\displaybreak[1]\\&
- \frac{235}{4} 
- \frac{65}{36}\*\pi^2
- \frac{7}{40}\*\pi^4
+ \frac{293}{18}\*\z3
\nonumber\displaybreak[1]\\&
+ \frac{\ep}{2}\*\left( -\frac{1497}{2} - \frac{601}{18}\*\pi^2 
     - \frac{511}{180}\*\pi^4 
     + \frac{1555}{9}\*\z3
     + \frac{181}{27}\*\pi^2\*\z3 + \frac{1813}{15}\*\z5\right)
\nonumber\displaybreak[1]\\&
- 3088.5\underline{5}385540997021022612585452885699551634307800797\*\ep^2 
\nonumber\displaybreak[1]\\&
- 14498.1092918745193546816622992228822221794484130674\*\ep^3 
\nonumber\displaybreak[1]\\&
- 75204.1336314891447627178739121420142533755186792735\*\ep^4 
\nonumber\displaybreak[1]\\&
- 315844.168768264796655463071293944512650159761067379\*\ep^5
%\nonumber\displaybreak[1]\\&
+\mathcal{O}(\ep^6)
\,,\displaybreak[1]\\%%%%%%%%%%%%%%%%%%%%%%%%%%%%%%%%%%%%%%%%%%%%%%%% 
\TopL{416017}&=
  \frac{7}{8\*\ep^4}
+ \frac{85}{16\*\ep^3}
+ \frac{1}{8\*\ep^2}\*\left(\frac{601}{4} + \frac{13}{3}\*\pi^2\right)
+ \frac{1}{\ep}\*\left(\frac{2747}{64} 
    + \frac{155}{48}\*\pi^2 
    - \frac{8}{3}\*\z3\right)
\nonumber\displaybreak[1]\\&
+ \frac{2329}{128}
+ \frac{1091}{96}\*\pi^2
+ \frac{\pi^4}{4}
+ \frac{43}{2}\*\z3
+ \frac{17}{2}\*\sqrt{3}\*\pb2
+ 2\*\pb2^2 
\nonumber\displaybreak[1]\\&
+ 6.4118\underline{7}248372530806306191468163270610970148366992008\*\ep 
\nonumber\displaybreak[1]\\&
- 1028.33128833387533834458667352622680634323915257341\*\ep^2 
\nonumber\displaybreak[1]\\&
- 16249.8777552641957747492917792334716880522254841241\*\ep^3 
\nonumber\displaybreak[1]\\&
- 65696.8989733875039196986140933337667282094540485106\*\ep^4 
\nonumber\displaybreak[1]\\&
- 430175.281094824821441568751585189206237721930681617\*\ep^5
%\nonumber\displaybreak[1]\\&
+\mathcal{O}(\ep^6)
\,,\displaybreak[1]\\%%%%%%%%%%%%%%%%%%%%%%%%%%%%%%%%%%%%%%%%%%%%%%%%
\TopL{416022}&= 
  \frac{5}{8\*\ep^4}
+ \frac{33}{8\*\ep^3}
+ \frac{1}{24\*\ep^2}\*\left(\frac{775}{2} + 11\*\pi^2\right)
\nonumber\displaybreak[1]\\&
+ \frac{1}{\ep}\*\left( \frac{1389}{32} + \frac{23}{8}\*\pi^2 
   + 3\*\sqrt{3}\*\pb2 - \frac{7}{3}\*\z3\right)
\nonumber\displaybreak[1]\\&
+ 147.807\underline{8}70353419457006020027881277752020796006112323
\nonumber\displaybreak[1]\\&
+ 467.918626785158952385003318864377551570797031294221\*\ep 
\nonumber\displaybreak[1]\\&
- 1988.53979380180044626102894892460888176325158879271\*\ep^2 
\nonumber\displaybreak[1]\\&
- 4066.05004959291604399936393129691406383824993650381\*\ep^3 
\nonumber\displaybreak[1]\\&
- 84210.5386674689601672799852884017698624685352637357\*\ep^4 
\nonumber\displaybreak[1]\\&
- 187546.771892814733657835744490067583363892471336553\*\ep^5
%\nonumber\displaybreak[1]\\&
+\mathcal{O}(\ep^6)
\,,\displaybreak[1]\\%%%%%%%%%%%%%%%%%%%%%%%%%%%%%%%%%%%%%%%%%%%%%%%%
\TopL{416021}&= 
  \frac{1}{2\*\ep^4}
+ \frac{47}{16\*\ep^3}
+ \frac{1}{\ep^2}\*\left(\frac{317}{32} + \frac{\pi^2}{3}\right)
%\nonumber\displaybreak[1]\\&
+ \frac{1}{\ep}\*\left(\frac{1315}{64} + \frac{31}{16}\*\pi^2 
   + 3\*\sqrt{3}\*\pb2 - \frac{5}{3}\*\z3\right)
\nonumber\displaybreak[1]\\&
+ 83.135\underline{8}882000807808465322734822592975462100986465128
\nonumber\displaybreak[1]\\&
+ 27.1740919045453848408661874005706354672639669874261\*\ep 
\nonumber\displaybreak[1]\\&
- 1554.56495046662560630454485042040757181079394589806\*\ep^2 
\nonumber\displaybreak[1]\\&
- 9710.65466554205477699849510014536535786834553811646\*\ep^3 
\nonumber\displaybreak[1]\\&
- 59651.1029754541134802977172410309014967516228078574\*\ep^4 
\nonumber\displaybreak[1]\\&
- 263782.712760320335062339831118979031401524684903280\*\ep^5
%\nonumber\displaybreak[1]\\&
+\mathcal{O}(\ep^6)
\,,\displaybreak[1]\\%%%%%%%%%%%%%%%%%%%%%%%%%%%%%%%%%%%%%%%%%%%%%%%%
\TopL{41603}&= 
  \frac{5}{24\*\ep^4}
+ \frac{55}{48\*\ep^3}
+ \frac{1}{24\*\ep^2}\*\left(\frac{331}{4} + \frac{11}{3}\*\pi^2\right)
\nonumber\displaybreak[1]\\& 
+ \frac{1}{3\*\ep}\*\left(\frac{977}{64} + \frac{121}{48}\*\pi^2 
    + \frac{17}{3}\*\z3\right)
- \frac{1831}{128}
+ \frac{661\*\pi^2}{288} 
- \frac{11\*\pi^4}{120}
+ \frac{187}{18}\*\z3
\nonumber\displaybreak[1]\\& 
- \ep\*\left( \frac{42485}{256} - \frac{335}{576}\*\pi^2
   + \frac{121}{240}\*\pi^4 
   - \frac{1327}{36}\*\z3
   - \frac{77}{27}\*\pi^2\*\z3 - \frac{172}{3}\*\z5\right)
\nonumber\displaybreak[1]\\& 
- 1308.40\underline{0}45975490723588094289794752583132729289225400\*\ep^2 
\nonumber\displaybreak[1]\\& 
- 6297.44410007372169525480920361163851686356380177551\*\ep^3 
\nonumber\displaybreak[1]\\& 
- 39381.3291846386139130702622754006165033230617765764\*\ep^4 
\nonumber\displaybreak[1]\\& 
- 156879.079031917198094441108322254183097437076390786\*\ep^5
%\nonumber\displaybreak[1]\\&
+\mathcal{O}(\ep^6)
\,,\displaybreak[1]\\%%%%%%%%%%%%%%%%%%%%%%%%%%%%%%%%%%%%%%%%%%%%%%%%
\TopL{417044}&=
  \frac{1}{8\*\ep^4}
+ \frac{5}{4\*\ep^3}
+ \frac{1}{\ep^2}\*\left(\frac{65}{8} + \frac{\pi^2}{8} + \z3\right)
\nonumber\displaybreak[1]\\& 
+ \frac{1}{\ep}\*\left(\frac{175}{4} + \frac{3}{4}\*\pi^2 
    + \frac{\pi^4}{60} 
    - 6\*\sqrt{3}\*\pb2 + \frac{10}{3}\*\z3\right)
\nonumber\displaybreak[1]\\& 
+ 188.248\underline{4}93841520722512481676436388871071949826368957
\nonumber\displaybreak[1]\\& 
+ 965.746295254204695422785736580849366762621253148187\*\ep 
\nonumber\displaybreak[1]\\& 
+ 2625.62618545640179663793065740741631915228108024570\*\ep^2 
\nonumber\displaybreak[1]\\& 
+ 17532.0585981528153344707844000089248687167371984949\*\ep^3 
\nonumber\displaybreak[1]\\& 
+ 36105.6708001198638661491047454223362925167216377188\*\ep^4 
\nonumber\displaybreak[1]\\& 
+ 300793.633336221637970252928259231319392170565232562\*\ep^5
%\nonumber\displaybreak[1]\\&
+\mathcal{O}(\ep^6)
\,,\displaybreak[1]\\%%%%%%%%%%%%%%%%%%%%%%%%%%%%%%%%%%%%%%%%%%%%%%%%
\TopL{417043}&=
  \frac{1}{12\*\ep^4}
+ \frac{5}{6\*\ep^3}
+ \frac{1}{\ep^2}\*\left(\frac{65}{12} + \frac{7}{36}\*\pi^2 + \z3\right)
\nonumber\displaybreak[1]\\& 
+ \frac{1}{3\*\ep}\*\left(\frac{175}{2} + \frac{23}{6}\*\pi^2 
   + \frac{\pi^4}{20} - \frac{13}{3}\*\z3\right)
\nonumber\displaybreak[1]\\& 
+ \frac{567}{4}
+ \frac{191}{36}\*\pi^2
+ \frac{19}{36}\*\pi^4
- \frac{367}{9}\*\z3
- \frac{1}{3}\pi^2\*\z3
- \frac{89}{3}\*\z5
+ 8\*\sqrt{3}\*\pb2
+ 4\*\pb2^2 
\nonumber\displaybreak[1]\\& 
+ 915.\underline{9}96742715562407821582232879837038545362469323820\*\ep 
\nonumber\displaybreak[1]\\& 
+ 2767.39379170194861041563852820827715025382193870924\*\ep^2 
\nonumber\displaybreak[1]\\& 
+ 16538.6792505904580566239289601519271328210126938973\*\ep^3 
\nonumber\displaybreak[1]\\& 
+ 40456.0909713394095410013351008764599207492777702247\*\ep^4 
\nonumber\displaybreak[1]\\& 
+ 280703.977577876240727702494488542000763547356431605\*\ep^5
%\nonumber\displaybreak[1]\\&
+\mathcal{O}(\ep^6)
\,,\displaybreak[1]\\%%%%%%%%%%%%%%%%%%%%%%%%%%%%%%%%%%%%%%%%%%%%%%%%
\TopL{417049}&=
  \frac{1}{8\*\ep^4}
+ \frac{13}{12\*\ep^3}
+ \frac{1}{\ep^2}\*\left(\frac{143}{24} + \frac{\pi^2}{8} + \z3\right)
%\nonumber\displaybreak[1]\\& 
+ \frac{1}{3\*\ep}\*\left(\frac{317}{4} + \frac{43}{12}\*\pi^2 
    + \frac{\pi^4}{20} + 19\*\z3\right)
\nonumber\displaybreak[1]\\& 
+ 158.265\underline{8}94985614754765609093664450928081621685254172
\nonumber\displaybreak[1]\\& 
+ 1040.51973958438238064321813618885698154945163332335\*\ep 
\nonumber\displaybreak[1]\\& 
+ 2220.52649830986620272733311300227293842057928920038\*\ep^2 
\nonumber\displaybreak[1]\\& 
+ 19009.3819367933781054485906380197274398245100237012\*\ep^3 
\nonumber\displaybreak[1]\\& 
+ 29872.2890554813361688482086053842835376892923973047\*\ep^4 
\nonumber\displaybreak[1]\\& 
+ 325486.618525642205718848428958393877946064518753281\*\ep^5
%\nonumber\displaybreak[1]\\&
+\mathcal{O}(\ep^6)
\,,\displaybreak[1]\\%%%%%%%%%%%%%%%%%%%%%%%%%%%%%%%%%%%%%%%%%%%%%%%%
\TopL{417051}&= 
  \frac{1}{6\*\ep^4}
+ \frac{3}{2\*\ep^3}
+ \frac{1}{\ep^2}\*\left(\frac{26}{3} + \frac{\pi^2}{18} 
   + \frac{\z3}{2}\right)
\nonumber\displaybreak[1]\\& 
+ \frac{1}{\ep}\*\left(41 + \frac{\pi^2}{2} 
       + \frac{\pi^4}{120} - \frac{7}{18}\*\z3\right)
\nonumber\displaybreak[1]\\& 
+ 192.14\underline{3}949985189792783133357258987262614727056122521
\nonumber\displaybreak[1]\\& 
+ 895.583232707441960151372330063515002337522803534682\*\ep 
\nonumber\displaybreak[1]\\& 
+ 3015.50609158685899428979534043459900694651565328232\*\ep^2 
\nonumber\displaybreak[1]\\& 
+ 15431.7141565054206165629974866237765016909275506281\*\ep^3 
\nonumber\displaybreak[1]\\& 
+ 45950.3851577110455999824698200183839583728323732333\*\ep^4 
\nonumber\displaybreak[1]\\& 
+ 256376.553671657904111860999953130364103694997971799\*\ep^5
%\nonumber\displaybreak[1]\\&
+\mathcal{O}(\ep^6)
\,,\displaybreak[1]\\%%%%%%%%%%%%%%%%%%%%%%%%%%%%%%%%%%%%%%%%%%%%%%%%
\TopL{417052}&= 
  \frac{1}{8\*\ep^4}
+ \frac{5}{4\*\ep^3}
+ \frac{1}{2\*\ep^2}\*\left(\frac{65}{4} + \frac{\pi^2}{4} 
      + \z3\right)
\nonumber\displaybreak[1]\\& 
+ \frac{1}{\ep}\*\left(\frac{175}{4} + \frac{3}{4}\*\pi^2 
     + \frac{\pi^4}{120} 
     - 6\*\sqrt{3}\*\pb2 - \frac{\z3}{6}\right)
\nonumber\displaybreak[1]\\& 
+ 210.851\underline{2}34758305984612090112605434985732961189359568
\nonumber\displaybreak[1]\\& 
+ 790.910352868677999468381955470293781862809895707932\*\ep 
\nonumber\displaybreak[1]\\& 
+ 3471.90742617992105936864664247656869003732285384889\*\ep^2 
\nonumber\displaybreak[1]\\& 
+ 13388.2008481237206516463529137761462371387930576955\*\ep^3 
\nonumber\displaybreak[1]\\& 
+ 54666.0539850126748720361226803316191993587586541377\*\ep^4 
\nonumber\displaybreak[1]\\& 
+ 219636.910846771617661122733716410439039757597225140\*\ep^5
%\nonumber\displaybreak[1]\\&
+\mathcal{O}(\ep^6)
\,,\displaybreak[1]\\%%%%%%%%%%%%%%%%%%%%%%%%%%%%%%%%%%%%%%%%%%%%%%%%
\TopL{417047}&= 
  \frac{1}{12\*\ep^4} 
+ \frac{5}{6\*\ep^3}
+ \frac{1}{2\*\ep^2}\*\left(\frac{65}{6} 
                  + \frac{7}{18}\*\pi^2 
                  + \z3\right)
\nonumber\displaybreak[1]\\& 
+ \frac{1}{6\*\ep}\*\left(175 + \frac{23\*\pi^2}{3}
                + \frac{\pi^4}{20} - \frac{89}{3}\*\z3\right)
\nonumber\displaybreak[1]\\& 
+ 194.097392\underline{6}55331|631239559043545
\nonumber\displaybreak[1]\\& 
 + 770.17891682309|6887339876558212\*\ep 
\nonumber\displaybreak[1]\\& 
 + 3399.9948243974|1716780023304896\*\ep^2
\nonumber\displaybreak[1]\\&
|+ 13282.1053505362611944372969993\*\ep^3
\nonumber\displaybreak[1]\\& 
|+ 54753.4641354130075585240962863\*\ep^4
+\mathcal{O}(\ep^5)
\,,\displaybreak[1]\\%%%%%%%%%%%%%%%%%%%%%%%%%%%%%%%%%%%%%%%%%%%%%%%%
\TopL{41704707x2}&=
  \frac{\z3}{2\*\ep^2}
+ \frac{1}{\ep}\*\left(\frac{\pi^4}{120} + \frac{3}{2}\*\z3\right)
\nonumber\displaybreak[1]\\& 
- 4.5950107\underline{7}9222|629663368197802
\nonumber\displaybreak[1]\\& 
+ 78.94876834312|825216261582947\*\ep
\nonumber\displaybreak[1]\\& 
- 340.518381398|9136598834071036\*\ep^2
\nonumber\displaybreak[1]\\& 
+ 1861.12230255|2258601194625502\*\ep^3
\nonumber\displaybreak[1]\\& 
|- 8297.001296673572071195678072\*\ep^4
+\mathcal{O}(\ep^5)
\,,\displaybreak[1]\\%%%%%%%%%%%%%%%%%%%%%%%%%%%%%%%%%%%%%%%%%%%%%%%%
\TopL{417048}&= 
  \frac{1}{12\*\ep^4}
+ \frac{2}{3\*\ep^3}
+ \frac{1}{\ep^2}\*\left(\frac{13}{4} + \frac{7}{36}\*\pi^2 + \z3\right)
\nonumber\displaybreak[1]\\& 
+ \frac{1}{3\*\ep}\*\left(\frac{71}{2} + \frac{14}{3}\*\pi^2 
            + \frac{\pi^4}{20} + \frac{68}{3}\*\z3\right)
\nonumber\displaybreak[1]\\& 
+ 154.60463\underline{1}|544617018499470855434
\nonumber\displaybreak[1]\\& 
|+ 894.807380418881266203622406485\*\ep
\nonumber\displaybreak[1]\\& 
|+ 2607.13652650580177242514674361\*\ep^2
%\nonumber\displaybreak[1]\\&
+\mathcal{O}(\ep^3)
\,,\displaybreak[1]\\%%%%%%%%%%%%%%%%%%%%%%%%%%%%%%%%%%%%%%%%%%%%%%%%
%%%%%%%%%%%%%%%%%%%%%%%%%%%%%%%%%%%%%%%%%%%%%%%%%%%
%%%%%%%%%%%%%%%%%%%%%%%%%%%%%%%%%%%%%%%%%%%%%%%%%%%
\TopL{41704804x2}&=
  \frac{1}{12\*\ep^4}
+ \frac{1}{3\*\ep^3}
+ \frac{7}{12\*\ep^2}\*\left( 1 + \frac{1}{3}\*\pi^2\right)
- \frac{1}{3\*\ep}\*\left( \frac{7}{2} - \frac{7}{3}\*\pi^2 
              - \frac{5}{3}\*\z3\right)
\nonumber\displaybreak[1]\\& 
+ 10.00336626|15671539152667896043
\nonumber\displaybreak[1]\\& 
|+ 121.389726250269120381127113931\*\ep
\nonumber\displaybreak[1]\\& 
|- 338.169587547191526007167931728\*\ep^2
%\nonumber\displaybreak[1]\\&
+\mathcal{O}(\ep^3)
\,,\displaybreak[1]\\%%%%%%%%%%%%%%%%%%%%%%%%%%%%%%%%%%%%%%%%%%%%%%%%
\TopL{417030}&=
  \frac{1}{6\*\ep^4}
+ \frac{3}{2\*\ep^3}
+ \frac{1}{3\*\ep^2}\*\left(26 + \frac{\pi^2}{6}\right)
+ \frac{1}{\ep}\*\left(41 + \frac{\pi^2}{2} - \frac{35}{9}\*\z3\right)
\nonumber\displaybreak[1]\\& 
+ \frac{1039}{6}
+ \frac{55}{18}\*\pi^2
+ \frac{37}{120}\*\pi^4
- \frac{94}{3}\*\z3
+ 2\*\sqrt{3}\*\pb2 
+ 10\*\pb2^2 
\nonumber\displaybreak[1]\\& 
+ 736.116\underline{8}32299866041378628875269512201227568815410747\*\ep 
\nonumber\displaybreak[1]\\& 
+ 3742.45984331167422662482544777380447491847387617756\*\ep^2 
\nonumber\displaybreak[1]\\& 
+ 11749.4302169340506083438761905399834852011067964612\*\ep^3 
\nonumber\displaybreak[1]\\& 
+ 62286.5836001930375873157912314697672224972850370815\*\ep^4 
\nonumber\displaybreak[1]\\& 
+ 184394.067587123000354685027809174508736973039378933\*\ep^5
%\nonumber\displaybreak[1]\\&
+\mathcal{O}(\ep^6)
\,,\displaybreak[1]\\%%%%%%%%%%%%%%%%%%%%%%%%%%%%%%%%%%%%%%%%%%%%%%%%
\TopL{417036}&= 
  \frac{1}{8\*\ep^4}
+ \frac{13}{12\*\ep^3}
+ \frac{1}{2\*\ep^2}\*\left(\frac{143}{12} + \frac{\pi^2}{4} + \z3\right)
% \nonumber\displaybreak[1]\\& 
+ \frac{1}{6\*\ep}\*\left(\frac{317}{2} + \frac{43}{6}\*\pi^2 
      + \frac{\pi^4}{20} + 17\*\z3\right)
 \nonumber\displaybreak[1]\\& 
+ \frac{2455}{24}
+ \frac{545}{72}\*\pi^2
+ \frac{\pi^4}{90}
+ \frac{169}{18}\*\z3
- \frac{\pi^2\*\z3}{6}
- \frac{59}{6}\*\z5
+ 2\*\sqrt{3}\*\pb2 
- 8\*\pb2^2 
\nonumber\displaybreak[1]\\&
+ 895.\underline{2}67223034507792336545934746971903090074177178563\*\ep 
\nonumber\displaybreak[1]\\& 
+ 2851.24487745129920226056389168977151286787950149531\*\ep^2 
\nonumber\displaybreak[1]\\& 
+ 15769.2924893036658768261052813143569858442259460015\*\ep^3 
\nonumber\displaybreak[1]\\& 
+ 44094.4294409565473747922022723147055746208297110835\*\ep^4 
\nonumber\displaybreak[1]\\& 
+ 262762.378299443371029842954614853325970041327609194\*\ep^5
%\nonumber\displaybreak[1]\\&
+\mathcal{O}(\ep^6)
\,,\displaybreak[1]\\%%%%%%%%%%%%%%%%%%%%%%%%%%%%%%%%%%%%%%%%%%%%%%%%
\TopL{417038}&= 
  \frac{2}{\ep^2}\*\z3
- \frac{1}{\ep}\*\left( \frac{\pi^4}{24} - 8\*\z3 + 6\*\pb2^2\right)
\nonumber\displaybreak[1]\\& 
+ 30.0908437\underline{8}21524694782206570213720645826480229016466
\nonumber\displaybreak[1]\\& 
- 36.8230691904535254355540493262162846080159401689486\*\ep 
\nonumber\displaybreak[1]\\& 
+ 260.507827461469390827561610096003346014392421090678\*\ep^2 
\nonumber\displaybreak[1]\\& 
- 562.809646826601965896917933562966364880244022357921\*\ep^3 
\nonumber\displaybreak[1]\\& 
+ 2222.67588565840309456690342209333550093550969179227\*\ep^4 
\nonumber\displaybreak[1]\\& 
- 6072.20564531822489618544102278797200412026771890767\*\ep^5
%\nonumber\displaybreak[1]\\&
+\mathcal{O}(\ep^6)
\,,\displaybreak[1]\\%%%%%%%%%%%%%%%%%%%%%%%%%%%%%%%%%%%%%%%%%%%%%%%%
\TopL{418040}&=
  \frac{5}{\ep}\*\z5
\nonumber\displaybreak[1]\\& 
- 20.858597\underline{6}739540823192137169923590025170233079442547
\nonumber\displaybreak[1]\\& 
+ 148.814327289184790880960752283863066723514605901922\*\ep 
\nonumber\displaybreak[1]\\& 
- 657.418920712168196526528768117001266054030349661525\*\ep^2 
\nonumber\displaybreak[1]\\& 
+ 3182.47221782449086620274231595579830546198641810013\*\ep^3 
\nonumber\displaybreak[1]\\& 
- 13795.8211363763697941825608377160301474614616561880\*\ep^4
%\nonumber\displaybreak[1]\\&
+\mathcal{O}(\ep^5)
\,,\displaybreak[1]\\%%%%%%%%%%%%%%%%%%%%%%%%%%%%%%%%%%%%%%%%%%%%%%%%
\TopL{418042}&=
  \frac{5}{\ep}\*\z5
\nonumber\displaybreak[1]\\&\phantom{=}
- 19.111865\underline{7}0594|90732311103013898
\nonumber\displaybreak[1]\\&\phantom{=}
+ 141.144175505|094858012998872298\*\ep
\nonumber\displaybreak[1]\\&\phantom{=}
- 605.43119057|5217055408942299967\*\ep^2
\nonumber\displaybreak[1]\\&\phantom{=}
|+ 2945.58041293149149762720437022\*\ep^3
+\mathcal{O}(\ep^4)
\,,\displaybreak[1]\\%%%%%%%%%%%%%%%%%%%%%%%%%%%%%%%%%%%%%%%%%%%%%%%%
\TopL{418044}&=
  \frac{5}{\ep}\*\z5
\nonumber\displaybreak[1]\\&
- 19.079375\underline{0}927079960314257405875623941525789104197843 
\nonumber\displaybreak[1]\\&
+ 141.252248186107747164092632797489850337070843975841\*\ep 
\nonumber\displaybreak[1]\\&
- 605.02962120187025890466267145520655064567103503240\*\ep^2 
\nonumber\displaybreak[1]\\&
+ 2946.48740435117339409606564510027475583646652772\*\ep^3 
\nonumber\displaybreak[1]\\&
- 12664.231982725689641172668688917148051149093738\*\ep^4 
\nonumber\displaybreak[1]\\&
+ 54825.71931364361072956312742419885158634286091\*\ep^5
%\nonumber\displaybreak[1]\\&
+\mathcal{O}(\ep^6)
\,,\displaybreak[1]\\%%%%%%%%%%%%%%%%%%%%%%%%%%%%%%%%%%%%%%%%%%%%%%%%
\TopL{418048}&=
  \frac{5}{\ep}\*\z5
- 18.7699726\underline{9}|85410818615637417243
\nonumber\displaybreak[1]\\&\phantom{=}
|+ 142.191941064595949485071666672\*\ep
\nonumber\displaybreak[1]\\&\phantom{=}
|- 600.255511127362803430335175869\*\ep^2
\nonumber\displaybreak[1]\\&\phantom{=}
|+ 2957.18558813594868083060620764\*\ep^3
+\mathcal{O}(\ep^4)
\,,\displaybreak[1]\\%%%%%%%%%%%%%%%%%%%%%%%%%%%%%%%%%%%%%%%%%%%%%%%%
\TopL{418036}&=
  \frac{5}{\ep}\*\z5
\nonumber\displaybreak[1]\\&
- 7.8587\underline{6}279518922242360822990783004542937882100674500
\nonumber\displaybreak[1]\\&
+ 128.531386676431303444223252316571121713818949997681\*\ep 
\nonumber\displaybreak[1]\\&
- 349.879176215675436895102723350659482216370181950816\*\ep^2 
\nonumber\displaybreak[1]\\&
+ 2332.83399433790756728143973285449385093610438970117\*\ep^3 
\nonumber\displaybreak[1]\\&
- 8346.39781820188664641354322284607837497614943630437\*\ep^4 
\nonumber\displaybreak[1]\\&
+ 40158.5604096624536390379365615442757755373577157863\*\ep^5
%\nonumber\displaybreak[1]\\&
+\mathcal{O}(\ep^6)
\,,\displaybreak[1]\\%%%%%%%%%%%%%%%%%%%%%%%%%%%%%%%%%%%%%%%%%%%%%%%%
\TopL{418037}&= 
  \frac{5}{\ep}\*\z5
\nonumber\displaybreak[1]\\&
- 13.72063\underline{0}2625920162070487342658557388782474967767706
\nonumber\displaybreak[1]\\&
+ 135.928906108952872038638377087143640185611677958787\*\ep 
\nonumber\displaybreak[1]\\&
- 497.764644552772376477970102393026516204883127610554\*\ep^2 
\nonumber\displaybreak[1]\\&
+ 2695.35125395194610632909729485139956434354033178137\*\ep^3 
\nonumber\displaybreak[1]\\&
- 11004.9478850481681338765147169019923471294985485170\*\ep^4 
\nonumber\displaybreak[1]\\&
+ 49153.9490476063842090788328486862743934773519776031\*\ep^5
%\nonumber\displaybreak[1]\\&
+\mathcal{O}(\ep^6)
\,,\displaybreak[1]\\%%%%%%%%%%%%%%%%%%%%%%%%%%%%%%%%%%%%%%%%%%%%%%%%
\TopL{418033}&=
  \frac{5}{\ep}\*\z5
\nonumber\displaybreak[1]\\&
+ 0.3748\underline{4}4191926851274910805165871097566997942271321280
\nonumber\displaybreak[1]\\&
+ 141.683133328263640451220557622143025514756763189200\*\ep 
\nonumber\displaybreak[1]\\&
- 146.684014785607112711629455842851698248300684175120\*\ep^2 
\nonumber\displaybreak[1]\\&
+ 2448.77867872444787260224256254958000162624856674133\*\ep^3 
\nonumber\displaybreak[1]\\&
- 5421.70587793103981619383560960628718332878155020120\*\ep^4 
\nonumber\displaybreak[1]\\&
+ 38306.6106494119866058980990395532741466206014427522\*\ep^5
%\nonumber\displaybreak[1]\\&
+\mathcal{O}(\ep^6)
\,,\displaybreak[1]\\%%%%%%%%%%%%%%%%%%%%%%%%%%%%%%%%%%%%%%%%%%%%%%%%
\TopL{418034}&=
  \frac{5}{\ep}\*\z5
\nonumber\displaybreak[1]\\&
+ 7.304459\underline{6}8508659934705130294982328536505673135057504
\nonumber\displaybreak[1]\\&
+ 141.443417856878944115040221346329269896865316839719\*\ep 
\nonumber\displaybreak[1]\\&
+ 40.4773793831579631350040651321790566085488080065966\*\ep^2 
\nonumber\displaybreak[1]\\&
+ 2244.39062716662154526526360631628245537948712203304\*\ep^3 
\nonumber\displaybreak[1]\\&
- 2188.78999525984793680680710077425046912668999350376\*\ep^4 
\nonumber\displaybreak[1]\\&
+ 31134.4838067073617857861615173845872142049443337933\*\ep^5
%\nonumber\displaybreak[1]\\&
+\mathcal{O}(\ep^6)
\,,\displaybreak[1]\\%%%%%%%%%%%%%%%%%%%%%%%%%%%%%%%%%%%%%%%%%%%%%%%%
\TopL{41807}&=
  \frac{3}{2\*\ep^2}\*\z3 
- \frac{1}{\ep}\*\left( \frac{\pi^4}{20} 
    - \frac{15}{2}\*\z3 + 6\*\pb2^2\right)
\nonumber\displaybreak[1]\\&
+ 40.477\underline{3}901150829247806609147361914210374485307884608
\nonumber\displaybreak[1]\\&
- 100.628977840966481693832556085825783093037821912010\*\ep 
\nonumber\displaybreak[1]\\&
+ 679.056339067891403282049708893797069643756722410073\*\ep^2 
\nonumber\displaybreak[1]\\&
- 2387.60452082758635632061839710152233583503731592210\*\ep^3 
\nonumber\displaybreak[1]\\&
+ 11259.2595068584556862116342453958677140933294272601\*\ep^4 
\nonumber\displaybreak[1]\\&
- 44413.9549636399070618153357715272931135284827413638\*\ep^5
%\nonumber\displaybreak[1]\\&
+\mathcal{O}(\ep^6)
\,,\displaybreak[1]\\%%%%%%%%%%%%%%%%%%%%%%%%%%%%%%%%%%%%%%%%%%%%%%%%
\TopL{419012}={}&
- 2.426956\underline{3}|9537700735
\nonumber\displaybreak[1]\\&
|+ 4.01554669961524192\*\ep
\nonumber\displaybreak[1]\\&
|- 43.6962533603324647\*\ep^2
\nonumber\displaybreak[1]\\&
|+ 128.936157875347890\*\ep^3
+\mathcal{O}(\ep^4)
\,,\displaybreak[1]\\%%%%%%%%%%%%%%%%%%%%%%%%%%%%%%%%%%%%%%%%%%%%%%%%
%%%%%%%%%%%%%%%%%%%%%%%%%%%%%%%%%%%%%%%%%%%%%%%%%%%
%%%%%%%%%%%%%%%%%%%%%%%%%%%%%%%%%%%%%%%%%%%%%%%%%%%
\TopL{41901205x2}={}&
+ 0.4736|11472272364450
\nonumber\displaybreak[1]\\&
|+ 1.09585342206826990\*\ep
\nonumber\displaybreak[1]\\&
|+ 5.37764333252884269\*\ep^2
\nonumber\displaybreak[1]\\&
|+ 8.82896457590640998\*\ep^3
+\mathcal{O}(\ep^4)
\,,\displaybreak[1]\\%%%%%%%%%%%%%%%%%%%%%%%%%%%%%%%%%%%%%%%%%%%%%%%%
\TopL{419011}={}&
- 3.7114026\underline{4}|536682392682628965373
\nonumber\displaybreak[1]\\&
|- 2.11520599545877264756135261804\*\ep
\nonumber\displaybreak[1]\\&
|- 71.9899451389829491830674629550\*\ep^2
\nonumber\displaybreak[1]\\&
|+ 41.1881294174309244417864419468\*\ep^3
+\mathcal{O}(\ep^4)
\,,\displaybreak[1]\\%%%%%%%%%%%%%%%%%%%%%%%%%%%%%%%%%%%%%%%%%%%%%%%%
\TopL{419010}={}&
- 6.7284\underline{7}05600856|8105547188977521
\nonumber\displaybreak[1]\\&
- 26.08764659996|66155389659770717\*\ep
\nonumber\displaybreak[1]\\&
- 214.6477179124|11362028052727052\*\ep^2
\nonumber\displaybreak[1]\\&
- 613.715203096|626075654874908838\*\ep^3
%\nonumber\displaybreak[1]\\&
+\mathcal{O}(\ep^4)
\,,\displaybreak[1]\\%%%%%%%%%%%%%%%%%%%%%%%%%%%%%%%%%%%%%%%%%%%%%%%%
\TopL{419017}={}&
- 3.449751\underline{1}|317390349922288
\nonumber\displaybreak[1]\\&
|+ 6.3127694885459812824115\*\ep
\nonumber\displaybreak[1]\\&
|- 63.668771344187502234181\*\ep^2
\nonumber\displaybreak[1]\\&
|+ 196.34402612627359322923\*\ep^3
+\mathcal{O}(\ep^4)
\,,%\displaybreak[1]\\%%%%%%%%%%%%%%%%%%%%%%%%%%%%%%%%%%%%%%%%%%%%%%%%
\end{align}
\noindent
with the symbol $\pb2$ defined by: 
\begin{equation}
\pb2=\mbox{Cl}_2\left(\frac{\pi}{3}\right)=\mbox{Im}\left[\mbox{Li}_{2}(e^{i\*\tfrac{\pi}{3}})\right]=1.0149416064096536250...\,.
\end{equation}
In ref.~\cite{Boughezal:2006xk} the set of master integrals has been
chosen slightly different. Instead of the topologies $T_{5,9}$ and
$T_{7,11}$ the topologies $T_{5,11}$ and $T_{7,17}$ have been
chosen as master integrals. These different master integrals can be
related to each other by the IBP-relations. For these integrals we also
give the analytical results, which we obtained for the leading
$\ep$-orders:  
\begin{align}
\TopL{4150401x2}&=
- \frac{1}{4\*\ep^4}
- \frac{9}{8\*\ep^3}
- \frac{1}{2\*\ep^2}\*\left( 5 + \frac{\pi^2}{6}\right)
+ \frac{1}{\ep}\*\left(\frac{15}{8} - \frac{3}{8}\*\pi^2  
         - 3\*\sqrt{3}\*\pb2 + \frac{\z3}{3}\right)
\nonumber\displaybreak[1]\\&\phantom{=}
+ 30.025\underline{3}835218317623192659472127405301377841888425075
\nonumber\displaybreak[1]\\&\phantom{=}
+ 254.111031964704093539375826206525134072256004057109\*\ep
\nonumber\displaybreak[1]\\&\phantom{=}
+ 1805.36593366425290012409359932346164787196465139299\*\ep^2
\nonumber\displaybreak[1]\\&\phantom{=}
+ 8898.24672362282959032538518672895187402701187699568\*\ep^3
\nonumber\displaybreak[1]\\&\phantom{=}
+ 43751.2551818001444717625115350236214456071856425684\*\ep^4
\nonumber\displaybreak[1]\\&\phantom{=}
+ 193367.012317936435653691364175466657628564760490824\*\ep^5
%\nonumber\displaybreak[1]\\&
+\mathcal{O}(\ep^6)
\,,\displaybreak[1]\\%%%%%%%%%%%%%%%%%%%%%%%%%%%%%%%%%%%%%%%%%%%%%%%%
\TopL{41704703x2}&=
- \frac{1}{8\*\ep^4}
- \frac{3}{4\*\ep^3}
- \frac{1}{8\*\ep^2}\*\left( 25 - \frac{\pi^2}{3}\right)
- \frac{1}{\ep}\*\left( \frac{45}{4} - \frac{7}{12}\*\pi^2 
              + \frac{4}{3}\*\z3\right)
\nonumber\displaybreak[1]\\&\phantom{=}
- \frac{301}{8}
+ \frac{27}{8}\*\pi^2
+ \frac{29}{720}\*\pi^4
- \frac{55}{3}\*\z3
- 4\*\sqrt{3}\*\pb2 
+ 7\*\pb2^2 
\nonumber\displaybreak[1]\\&\phantom{=}
- 17.7874\underline{4}259404|76975226672333232\*\ep
\nonumber\displaybreak[1]\\&\phantom{=}
- 175.317651447|398940503032885592\*\ep^2
\nonumber\displaybreak[1]\\&\phantom{=}
+ 313.98310535|3518264983361554717\*\ep^3
\nonumber\displaybreak[1]\\&\phantom{=}
|- 2142.79466723532874375769322841\*\ep^4
%\nonumber\displaybreak[1]\\&
+\mathcal{O}(\ep^5)
\label{41704703x2}\,.%\displaybreak[1]\\%%%%%%%%%%%%%%%%%%%%%%%%%%%%%%%%%%%%%%%%%%%%%%%%
\end{align}
All results are in full agreement with those published in
\cite{Boughezal:2006xk}.\\ 
An overview over the different members of the two sets of master
integrals discussed in section~\ref{sec:epf} and \ref{sec:standard} is 
given in table~\ref{tab:masterlist}.\\
\begin{table}[!ht]
\begin{center}
\begin{tabular}{|p{23ex}|l@{}r@{}*{7}{p{3.5ex}}p{0ex}|}\hline
\mbox{Set of master} \mbox{integrals, which are} \mbox{equal in both basis} & $M_{eql}$ & $ = \{$ 
& $T_{6,11}$, $T_{7,13}$, $T_{8, 6}$, $T_{9, 5}$, 
& $T_{7, 5}$, $T_{7,16}$, $T_{8, 7}$, $T_{9, 6}$, 
& $T_{7, 7}$, $T_{8, 1}$, $T_{8, 8}$, $T_{9, 7}$\} 
& $T_{7, 8}$, $T_{8, 2}$, $T_{8, 9}$, 
& $T_{7, 9}$, $T_{8, 3}$, $T_{8,10}$, 
& $T_{7,10}$, $T_{8, 4}$, $T_{9, 1}$, 
& $T_{7,12}$, $T_{8, 5}$, $T_{9, 3}$, & \\ \hline
\mbox{Set of additional} \mbox{members of the} \mbox{standard basis} & $M_{std} $ & $ = \{$ 
& $T_{5, 2}$, $T_{6, 4}$, $T_{6,16}$, $T_{7,11}$, 
& $T_{5, 4}$, $T_{6, 9}$, $T_{6,17}$, $T_{7,14}$, 
& $T_{5, 8}$, $T_{6,10}$, $T_{6,18}$, $T_{7,15}$, 
& $T_{5, 9}$, $T_{6,12}$, $T_{6,19}$, $T_{9, 4}$\} 
& $T_{5,10}$, $T_{6,13}$, $T_{7, 1}$, 
& $T_{6, 1}$, $T_{6,14}$, $T_{7, 2}$, 
& $T_{6, 2}$, $T_{6,15}$, $T_{7, 6}$, & \\ \hline
\mbox{Set of additional} \mbox{members of the} \mbox{$\ep$-finite basis} & $M_{epf} $ & $ = \{$ 
& $T^f_{5, 2}$, $T^f_{6, 4}$, $T^f_{6,16}$, $T^f_{7,11}$, 
& $T^f_{5, 4}$, $T^f_{6, 9}$, $T^f_{6,17}$, $T^f_{7,14}$, 
& $T^f_{5, 8}$, $T^f_{6,10}$, $T^f_{6,18}$, $T^f_{7,15}$, 
& $T^f_{5, 9}$, $T^f_{6,12}$, $T^f_{6,19}$, $T^f_{9, 4}$\} 
& $T^f_{5,10}$, $T^f_{6,13}$, $T^f_{7, 1}$, 
& $T^f_{6, 1}$, $T^f_{6,14}$, $T^f_{7, 2}$, 
& $T^f_{6, 2}$, $T^f_{6,15}$, $T^f_{7, 6}$, & \\ \hline
\mbox{Set of master integrals,} \mbox{known analytically to} \mbox{sufficient orders in $\ep$} & $M_{ana} $ & $ = \{$ 
& $T_{4, 1}$, $T_{6, 5}$, 
& $T_{5, 1}$, $T_{6, 6}$, 
& $T_{5, 3}$, $T_{6, 7}$, 
& $T_{5, 5}$, $T_{6, 8}$, 
& $T_{5, 6}$, $T_{7, 3}$, 
& $T_{5, 7}$, $T_{7, 4}$, 
& $T_{6, 3}$, $T_{9, 2}$\} & \\ \hline
\end{tabular}
\end{center}
\caption{\label{tab:masterlist} The standard basis is given by
  $M_{eql}\cup M_{std}\cup M_{ana}$.  The $\ep$-finite basis is given by
  $M_{eql}\cup M_{epf}\cup M_{ana}$. Note that the simple master
  integrals in the set $M_{ana}$ have not been replaced in the case of
  the existence of a spurious pole in the coefficient function, since
  these master integrals are known analytically to high orders in
  $\ep$. The master integrals of the table, which have not been given in
  this work can be found in ref.~\cite{Schroder:2005va,Chetyrkin:2006dh}.}
\end{table}\\
Finally, we want to make a  few comments about the results for the master
integrals in the standard basis as collected in
eqs. (\ref{41504})-(\ref{41704703x2}).
\begin{itemize}
\item For the calculation of the $\rho$-parameter the coefficients of
the master integrals shown in eqs.~(\ref{41504})-(\ref{41704703x2})
suffer from not more than a $1/\ep^2$-pole. On general grounds it is
clear that in any foreseeable {\em four}-loop calculation one would
always need the analytical terms, displayed in eqs.
(\ref{41504})-(\ref{41704703x2}) of the $\ep$-expansion of a master
integral and additionally up to at most the order $\ep^2$ for the
$\rho$-parameter.
Thus all other higher order $\ep$-terms, shown in eqs.
(\ref{41504})-(\ref{41704703x2}), will become physically relevant only
for five and more loops.

\item The use of the $\ep$-finite basis allows for an easy and
automatically {\em analytical} determination of a ``trivial'' part (that
is eventually expressible through the three-loop tadpoles) of the
$\ep$-expansion of a four-loop tadpole master
integral\cite{Chetyrkin:2006dh}.  This is just enough to control the
cancellation of the UV poles during the renormalization procedure in
completely algebraic way. In order to have the final physical result
completely analytically one needs to know at most one more term in the
$\ep$-expansion of the members of the $\ep$-finite basis in an
analytical form. In relatively simple cases this could be achieved with
analytical summation of auxiliary series constructed with the help of
either the differential relations
\cite{Kotikov:1990kg,Kotikov:1991hm,Kotikov:1991pm,Remiddi:1997ny} or
the difference ones
\cite{Laporta:1996mq,Laporta:2000dc,Laporta:2001dd,Laporta:2001rc,Laporta:2002pg}.
For recent progress in this direction see
e.g. \cite{Kalmykov:2005hb,Kniehl:2005yc,Kalmykov:2006pu,Bejdakic:2006vg}
and references therein.
\item One should, however, keep in mind that the numerical accuracy of
the Laporta method of computing master integrals is huge (30-50
digits!). This means that the availability of a completely analytical
result is absolutely irrelevant for phenomenology.
\end{itemize}

\section{Conclusions\label{sec:conc}}

We have constructed an $\ep$-finite basis for problems related to the
$\rho$-parameter. The calculation of the elements of this basis via the
Pad\'e method allows to obtain a lot of exact analytical relations among
various terms of the Taylor expansion in $\ep$ of the master
integrals. In particular we found analytically the pole parts of the
master integrals in both bases, the standard and the $\ep$-finite
one. For some members of the standard basis analytical results are also
obtained for orders beyond the pole part. With the help of the
$\ep$-finite basis the cancellation of all divergences during the
renormalization procedure can be done analytically.  These relations are
indispensable for any attempts to provide, say, the four-loop QCD
contributions to the $\rho$ parameter in a completely analytical form.
They also serve as a powerful check of available numerical results.  We
find full agreement with the results of the standard basis also
calculated in ref.~\cite{Boughezal:2006xk}. The conversion between the
two descriptions as well as all the results presented in this work will
be made available in computer readable form under the URL {\tt
http://www-ttp.physik.uni-karlsruhe.de/Progdata/ttp06/ttp06-30}.

\subsection*{Acknowledgments}

The authors are grateful to K.G.~Chetyrkin and J.H.~K{\"u}hn for
inspiring discussions as well as to M.~Czakon, Y. Schr{\"o}der and
M.~Steinhauser for a careful reading of the manuscript. The work was
supported by the Deutsche Forschungsgemeinschaft in the
Sonderforschungsbereich/Transregio SFB/TR-9 ``Computational Particle
Physics''.  The work of C.S. was also partially supported by MIUR under
contract 2001023713$\_$006.

\appendix
\section{Master integrals excluded in the construction\\ 
         of the $\boldsymbol{\ep}$-finite basis \label{app:analytic}}
In this appendix we give for completeness the result of the integrals,
which have been excluded in the construction of an $\ep$-finite basis,
since they are known analytically to higher powers in the
$\ep$-expansion. The results read:
\begin{align}
\TopF{41501}
&=
    3\*\Exp{4\*\ep\*\Egamma}\*
    \PGamma{5-2\*d}{}\*
    \PGamma{3-\tfrac{3\*d}{2}}{}\*
    \PGamma{\tfrac{d}{2}-1}{3}
\label{41051}\,,\displaybreak[1]\\%%%%%%%%%%%%%%%%%%%%%%%%%%%%%%%%%%%%%%%%%%%%%%%%
\TopF{41502}
&=
    -3\*\Exp{4\*\ep\*\Egamma}\*(d-4)\*
    \PGamma{2 - d}{}\*
    \PGamma{3 - \tfrac{3}{2}\*d}{}\*
    \PGamma{3 - \tfrac{d}{2}}{}\*
    \PGamma{\tfrac{d}{2}-2}{2}
\,,\displaybreak[1]\\%%%%%%%%%%%%%%%%%%%%%%%%%%%%%%%%%%%%%%%%%%%%%%%%
\TopF{41503}
&=
   -\frac{32\*\Exp{4\*\ep\*\Egamma}\*
    \PGamma{2 - d}{}\*
    \PGamma{3 - \frac{d}{2}}{3}\*
    \PGamma{\frac{d}{2}-2}{}}
    {\left(d-4\right)^2\*\left(d-2\right)^2}
\,,\displaybreak[1]\\%%%%%%%%%%%%%%%%%%%%%%%%%%%%%%%%%%%%%%%%%%%%%%%%
\TopF{416010} 
&=
 -\frac{8\*\Exp{4\*\ep\*\Egamma}\*
  \PGamma{6-2\*d}{}\*
  \PGamma{6-\tfrac{3\*d}{2}}{}\*
  \PGamma{3-\tfrac{d}{2}}{2}\*
  \PGamma{\tfrac{d}{2}-1}{4}\*
  \PGamma{\tfrac{3\*d}{2}-5}{}}
  {(d-4)^2\*(d-2)\*
  \PGamma{4-d}{}\*
  \PGamma{d-2}{2}} 
\,,\displaybreak[1]\\%%%%%%%%%%%%%%%%%%%%%%%%%%%%%%%%%%%%%%%%%%%%%%%%
\TopF{416018} 
&=
  \frac{
  12\*\Exp{4\*\ep\*\Egamma}\*
  \PGamma{8-2\*d}{}\*
  \PGamma{4-d}{}\*
  \PGamma{3-\tfrac{d}{2}}{}\*
  \PGamma{2\*d-8}{}\*
  \PGamma{\tfrac{d}{2}-1}{4}}
 {(d-3)\*
  \PGamma{d-2}{}\*
  \PGamma{\tfrac{3\*d}{2}-2}{}} 
\,,\displaybreak[1]\\%%%%%%%%%%%%%%%%%%%%%%%%%%%%%%%%%%%%%%%%%%%%%%%%
\TopF{41604}
&=
  -\frac{
  256\*\Exp{4\*\ep\*\Egamma}\*
  \PGamma{6 - \tfrac{3}{2}\*d}{}\*
  \PGamma{\tfrac{d}{2}}{3}\*
  \PGamma{3 - \tfrac{d}{2}}{3}\*
  \PGamma{\tfrac{3}{2}\*d-5}{}}
 {(d-4)^3\*(d-2)^3\*
  \PGamma{d-1}{2}}
\,,\displaybreak[1]\\%%%%%%%%%%%%%%%%%%%%%%%%%%%%%%%%%%%%%%%%%%%%%%%%
\TopF{41605}
&=
   4\*\Exp{4\*\ep\*\Egamma}\*
   \PGamma{2 - d}{2}\*
   \PGamma{3 - \tfrac{d}{2}}{2}\*
   \PGamma{\tfrac{d}{2}-2}{2}
\,,\displaybreak[1]\\%%%%%%%%%%%%%%%%%%%%%%%%%%%%%%%%%%%%%%%%%%%%%%%%
\TopF{417019} 
&=
 -\frac{
   32\*\Exp{4\*\ep\*\Egamma}\*
  \PGamma{8-2\*d}{}\*
  \PGamma{3-\frac{d}{2}}{3}\*
  \PGamma{2\*d-8}{}\*
  \PGamma{\frac{d}{2}-1}{5}}
  {(d-4)^2\*(d-2)\*
  \PGamma{d-2}{3}} 
\label{417019}\,,\displaybreak[1]\\%%%%%%%%%%%%%%%%%%%%%%%%%%%%%%%%%%%%%%%%%%%%%%%%
%%%%%%%%%%%%%%%%%%%%%%%%%%%%%%%%%%%%%%%%%%%%%%%%%%%%%%%%%%%%%%%%%%%%%
%%%%%%%%%%%%%%%%%%%%%%%%%%%%%%%%%%%%%%%%%%%%%%%%%%%%%%%%%%%%%%%%%%%%%
%%%%%%%%%%%%%%%%%%%%%%%%%%%%%%%%%%%%%%%%%%%%%%%%%%%%%%%%%%%%%%%%%%%%%
\TopL{417039} 
&=
\frac{1}{12\*\ep^4}
+ \frac{2}{3\*\ep^3}
+ \frac{1}{36\*\ep^2}\*\left(117 + 7\*\pi^2\right)
+ \frac{1}{18\*\ep}\*\left(213 + 28\*\pi^2 + 10\*\z3\right)
\nonumber\displaybreak[1]\\
&+ \frac{1}{180}\*\left(5655 + 1365\*\pi^2 + 64\*\pi^4 + 800\*\z3\right)
\nonumber\displaybreak[1]\\
&+ \frac{\ep}{270}\*
  \left(8910 + 7455\*\pi^2 + 768\*\pi^4 + 7470\*\z3 
       + 350\*\pi^2\*\z3 + 8082\*\z5\right)
\nonumber\displaybreak[1]\\
&+ \frac{\ep^2}{1620}\*
  \Big(-480195 + 118755\*\pi^2 + 22626\*\pi^4 + 1093\*\pi^6 + 273600\*\z3 
\nonumber\displaybreak[1]\\
  &\qquad\qquad+ 16800\*\pi^2\*\z3 - 40740\*\z3^2 + 387936\*\z5\Big)
+ \mathcal{O}(\ep^3)
\label{417039}\,,\displaybreak[1]\\%%%%%%%%%%%%%%%%%%%%%%%%%%%%%%%%%%%%%%%%%%%%%%%%
\TopL{41908}  
&=
  \frac{5\*\z5}{\ep} 
+ \frac{1}{378}\*\left(5\*\pi^6 + 6426\*\z3^2 + 5670\*\z5\right)
+ \frac{\ep}{630}\*
  \Big(25\*\pi^6 + 357\*\pi^4\*\z3 
\nonumber\displaybreak[1]\\
&+ 32130\*\z3^2 + 22050\*\z5 + 7350\*\pi^2\*\z5 +70875\*\z7\Big)
+ \mathcal{O}(\ep^2)\,.
\label{41908}
\end{align}
The integrals (\ref{41051})-(\ref{417019}) are easily found via a repeated
application of well-known analytical formulas for a one-loop massive
tadpole and a one-loop massless propagator. A less simple diagram
(\ref{417039}) can be extracted from
\cite{Broadhurst:1986bx,Kazakov:1983ns,Bierenbaum:2003ud} while the most
complicated non-planar one (\ref{41908}) from \cite{chet:prep,Bekavac:2005xs}.
%%%%%%%%%%%%%%%%%%%%%%%%%%%
%%
%%
%%
%%
\section{Relations among particular orders of \mbox{different} master integrals
  of the standard basis \label{app:SpecialRelations}}
In this appendix we present the relations among particular orders of
different master integrals as discussed in section~\ref{sec:standard}:
{
\allowdisplaybreaks
\begin{eqnarray}
%  1
&&T_{6,13}^{(0)}-T_{6,18}^{(0)}=-\frac{11561}{128}-\frac{1685\*\pi^2}{288}-\frac{\pi^4}{4}+\frac{9\*\sqrt{3}\*s_2}{2}+\frac{166\*\z3}{9},\\[\relb]
%  2
&&T_{7,10}^{(0)}-T_{7,11}^{(0)}=\frac{567}{4}+\frac{179\*\pi^2}{36}+\frac{187\*\pi^4}{360}+4\*\sqrt{3}\*s_2+2\*s_2^2-\frac{388\*\z3}{9},\\[\relb]
%  3
&&2\*T_{6,13}^{(0)}-3\*T_{6,14}^{(0)}=\frac{89}{4}+\frac{31\*\pi^2}{36}+\frac{13\*\pi^4}{60}+12\*\sqrt{3}\*s_2+6\*s_2^2-\frac{131\*\z3}{18},\\[\relb]
%  4
&&2\*T_{5,10}^{(0)}-3\*T_{6,10}^{(0)}=\frac{3651}{16}+5\*\pi^2+\frac{11\*\pi^4}{40}-9\*\sqrt{3}\*s_2+12\*s_2^2-\frac{379\*\z3}{6},\\[\relb]
%  5
&&2\*T_{5,10}^{(0)}-T_{6,9}^{(0)}=\frac{1731}{16}+\frac{13\*\pi^2}{9}+\frac{11\*\pi^4}{30}-19\*\sqrt{3}\*s_2+16\*s_2^2-\frac{211\*\z3}{6},\\[\relb]
%  6
&&6\*T_{5,9}^{(1)}+T_{5,10}^{(0)}+T_{6,11}^{(0)}=\nonumber\\*[\rela]
&&\qquad\frac{2097}{64}+\frac{241\*\pi^2}{48}-\frac{35\*\pi^4}{72}+20\*\sqrt{3}\*s_2-20\*s_2^2-\frac{\z3}{9},\\[\relb]
%  7
&&6\*T_{5,9}^{(1)}-T_{5,10}^{(0)}+3\*T_{6,12}^{(0)}=\nonumber\\*[\rela]
&&\qquad-\frac{5435}{64}-\frac{247\*\pi^2}{48}-\frac{\pi^4}{2}+9\*\sqrt{3}\*s_2-24\*s_2^2+\frac{43\*\z3}{6},\\[\relb]
%  8
&&4\*T_{6,13}^{(0)}-2\*T_{6,13}^{(1)}+3\*T_{6,14}^{(1)}-T_{8,10}^{(0)}=\nonumber\\*[\rela]
&&\qquad-\frac{323}{4}+\frac{71\*\pi^2}{12}-\frac{19\*\pi^4}{36}+\frac{95\*\z3}{9}+\frac{86\*\pi^2\*\z3}{27}+\frac{328\*\z5}{5},\\[\relb]
%  9
&&T_{5,10}^{(0)}+T_{6,17}^{(0)}+T_{7,5}^{(0)}=\nonumber\\*[\rela]
&&\qquad\frac{50447}{192}+\frac{2671\*\pi^2}{144}+\frac{71\*\pi^4}{180}-9\*\sqrt{3}\*s_2-\frac{45\*\z3}{2}+\frac{\pi^2\*\z3}{9}-\frac{437\*\z5}{9},\\[\relb]
% 10
&&5\*T_{5,8}^{(1)}-21\*T_{5,9}^{(1)}-2\*T_{7,14}^{(1)}=\nonumber\\*[\rela]
&&\qquad-\frac{2716223}{2592}-\frac{3533\*\pi^2}{216}-\frac{293\*\pi^4}{180}-\frac{467\*s_2}{\sqrt{3}}\nonumber\\*[\rela]
&&\qquad-\frac{4\*s_2^2}{3}+\frac{2585\*\z3}{9}+\frac{46\*\pi^2\*\z3}{27}+\frac{263\*\z5}{5},\\[\relb]
% 11
&&T_{5,8}^{(1)}-9\*T_{5,9}^{(1)}-2\*T_{7,10}^{(1)}+2\*T_{7,11}^{(1)}=\nonumber\\*[\rela]
&&\qquad-\frac{3471265}{2592}-\frac{4543\*\pi^2}{216}-\frac{229\*\pi^4}{45}-\frac{217\*s_2}{\sqrt{3}}\nonumber\\*[\rela]
&&\qquad+\frac{40\*s_2^2}{3}+\frac{893\*\z3}{3}-\frac{28\*\pi^2\*\z3}{27}+\frac{2133\*\z5}{5},\\[\relb]
% 12
&&T_{5,8}^{(1)}-15\*T_{5,9}^{(1)}-2\*T_{6,16}^{(1)}=\nonumber\\*[\rela]
&&\qquad\frac{8415131}{10368}-\frac{39595\*\pi^2}{864}+\frac{1039\*\pi^4}{180}-\frac{349\*s_2}{\sqrt{3}}+\frac{298\*s_2^2}{3}+\frac{64\*\pi^2\*\log^2(2)}{3}\nonumber\\*[\rela]
&&\qquad-\frac{64\*\log^4(2)}{3}-512\*\pa-\frac{11135\*\z3}{18}+\frac{10\*\pi^2\*\z3}{3}+\frac{51\*\z5}{5},\\[\relb]
% 13
&&T_{5,8}^{(1)}+15\*T_{5,9}^{(1)}+2\*T_{5,10}^{(0)}+4\*T_{5,10}^{(1)}+2\*T_{6,11}^{(1)}-6\*T_{6,12}^{(1)}+2\*T_{7,16}^{(0)}=\nonumber\\*[\rela]
&&\qquad\frac{957005}{2592}+\frac{11027\*\pi^2}{216}-\frac{23\*\pi^4}{30}+\frac{53\*s_2}{\sqrt{3}}-\frac{140\*s_2^2}{3}\nonumber\\*[\rela]
&&\qquad-\frac{238\*\z3}{9}-\frac{101\*\pi^2\*\z3}{27}+\frac{974\*\z5}{15},\\[\relb]
% 14
&&T_{5,8}^{(1)}+15\*T_{5,9}^{(1)}-2\*T_{5,10}^{(0)}+4\*T_{5,10}^{(1)}+2\*T_{6,11}^{(1)}-6\*T_{6,12}^{(1)}+12\*T_{6,13}^{(0)}-4\*T_{6,13}^{(1)}\nonumber\\*[\rela]
&&\qquad+6\*T_{6,14}^{(1)}+6\*T_{7,8}^{(0)}=\frac{2290913}{2592}+\frac{14459\*\pi^2}{216}-\frac{79\*\pi^4}{45}+\frac{179\*s_2}{\sqrt{3}}\nonumber\\*[\rela]
&&\qquad-\frac{212\*s_2^2}{3}+\frac{100\*\z3}{9}+\frac{53\*\pi^2\*\z3}{27}+\frac{634\*\z5}{5},\\[\relb]
% 15
&&4\*T_{5,8}^{(1)}-12\*T_{5,9}^{(1)}-5\*T_{5,10}^{(0)}-2\*T_{5,10}^{(1)}-T_{6,11}^{(1)}+3\*T_{6,12}^{(1)}+12\*T_{6,13}^{(0)}\nonumber\\*[\rela]
&&\qquad-4\*T_{6,13}^{(1)}+6\*T_{6,14}^{(1)}-3\*T_{7,9}^{(0)}=-\frac{964973}{1296}-\frac{6613\*\pi^2}{216}-\frac{43\*\pi^4}{72}\nonumber\\*[\rela]
&&\qquad-\frac{184\*s_2}{\sqrt{3}}+\frac{52\*s_2^2}{3}+\frac{1618\*\z3}{9}+\frac{29\*\pi^2\*\z3}{27}+16\*\z5,\\[\relb]
% 16
&&T_{5,8}^{(1)}+15\*T_{5,9}^{(1)}+T_{5,10}^{(0)}+4\*T_{5,10}^{(1)}+2\*T_{6,11}^{(1)}-6\*T_{6,12}^{(1)}+6\*T_{6,13}^{(0)}-4\*T_{6,13}^{(1)}\nonumber\\*[\rela]
&&\qquad+6\*T_{6,14}^{(1)}+3\*T_{6,17}^{(0)}-3\*T_{7,7}^{(0)}=\frac{1284721}{5184}+\frac{29521\*\pi^2}{432}-\frac{313\*\pi^4}{180}\nonumber\\*[\rela]
&&\qquad+\frac{8\*s_2}{\sqrt{3}}-\frac{176\*s_2^2}{3}-\frac{2551\*\z3}{18}+\frac{35\*\pi^2\*\z3}{27}+\frac{5372\*\z5}{15},\\[\relb]
% 17
&&7\*T_{5,8}^{(1)}-31\*T_{5,9}^{(1)}-3\*T_{5,10}^{(0)}-2\*T_{5,10}^{(1)}-T_{6,11}^{(1)}+3\*T_{6,12}^{(1)}+20\*T_{6,13}^{(0)}-8\*T_{6,13}^{(1)}\nonumber\\*[\rela]
&&\qquad+12\*T_{6,14}^{(1)}+2\*T_{9,3}^{(0)}+6\*T_{9,4}^{(0)}=-\frac{590641}{2592}+\frac{2627\*\pi^2}{216}+\frac{707\*\pi^4}{360}\nonumber\\*[\rela]
&&\qquad-\frac{508\*s_2}{\sqrt{3}}+\frac{376\*s_2^2}{3}+\frac{53\*\pi^4\*\log(2)}{15}+\frac{4\*\pi^2\*\log^2(2)}{3}-\frac{4\*\log^4(2)}{3}\nonumber\\*[\rela]
&&\qquad+\frac{16\*\pi^2\*\log^3(2)}{3}-\frac{16\*\log^5(2)}{5}-32\*\pa\nonumber\\*[\rela]
&&\qquad+384\*\pc+\frac{1817\*\z3}{18}+\frac{47\*\pi^2\*\z3}{18}-\frac{2127\*\z5}{5}.
\end{eqnarray}
}

\end{document}